\documentclass[draftclsnofoot, onecolumn, 12pt, journal]{IEEEtran} %
\pdfoutput=1
\usepackage{graphicx}%[pdftex]
\usepackage{subfigure}
\usepackage{epstopdf}
\usepackage{epsfig}
\usepackage{float}
\usepackage{array}
\usepackage{cite}
\usepackage{bm}
\usepackage{amssymb}
\usepackage{amsmath}
\usepackage{amsfonts}
\usepackage{algorithmic}
\usepackage{stfloats}

\usepackage{fancyhdr}

\makeatletter
\renewcommand{\maketag@@@}[1]{\hbox{\m@th\normalsize\normalfont#1}}%
\makeatother

\usepackage[ruled]{algorithm2e}

\usepackage{xcolor}

\graphicspath{{figure/}}


\begin{document}

\title{OFDM-Based Massive Connectivity for LEO Satellite Internet of Things}
\author{Yong Zuo, Mingyang Yue, Mingchen Zhang, Sixian Li,\\ Shaojie Ni, and Xiaojun Yuan

\thanks{
     Yong Zuo and Shaojie Ni are with the College of Electronic Science and Technology, National University of Defense Technology, Changsha, China (e-mail: zuoyong@nudt.edu.cn; nishaojie123@126.com).
     
	 Mingyang Yue, Mingchen Zhang, Sixian Li and Xiaojun Yuan are with the National Laboratory of Science and Technology on Communications, University of Electronic Science and Technology of China, Chengdu 611731, China (e-mail: \{myyue, zhangmingchen,sxli\}@std.uestc.edu.cn;  xjyuan@uestc.edu.cn).}
}

\maketitle

\thispagestyle{fancy}

\lfoot{This work has been submitted to the IEEE for possible publication.  Copyright may be transferred without notice, after which this version may no longer be accessible.}

\cfoot{}

\renewcommand{\headrulewidth}{0mm}

\begin{abstract}
Low earth orbit (LEO) satellite has been considered as a potential supplement for the terrestrial Internet of Things (IoT). In this paper, we consider grant-free non-orthogonal random access (GF-NORA) in orthogonal frequency division multiplexing (OFDM) system to increase access capacity and reduce access latency  for LEO satellite-IoT. We  focus on the joint device activity detection (DAD) and channel estimation (CE) problem at the satellite access point. 
 The delay and the Doppler effect of the LEO satellite channel are assumed to be partially compensated.  We propose an OFDM-symbol repetition technique to  better distinguish the residual Doppler frequency shifts, and 
 present a grid-based parametric probability  model to characterize channel sparsity in the delay-Doppler-user domain, as well as to characterize the relationship between the channel states and the device activity. Based on that, 
we develop a robust Bayesian message passing algorithm named modified variance state propagation (MVSP) for joint DAD and CE. 
 Moreover, to tackle the mismatch between the real channel and its on-grid representation, an expectation–maximization (EM) framework is proposed to learn the grid parameters. Simulation results demonstrate that our proposed algorithms significantly outperform the existing approaches in both activity detection probability and channel estimation accuracy.

\end{abstract}


\IEEEpeerreviewmaketitle

\section{Introduction}

In recent years,  the Internet of Things (IoT) has attracted much attention, and is expected to  support different applications such as  smart homes, smart gateways, environmental monitoring, and smart cities \cite{iot1, iot2}. Different from human-type communications, the IoT spawns  a new communication  scenario named massive connectivity,  where a huge number of devices can access a single base station (BS) in a sporadic manner. In this scenario, the conventional grant-based random access solutions designed for human-type communications becomes very inefficient due to severe latency caused by device collisions \cite{GF}.

Grant-free non-orthogonal random access (GF-NORA) has been considered as a promising technique for massive connectivity \cite{GF,iotgf}, where active devices are allowed to directly transmit pilots and data to the BS without waiting for a grant. As such, in each transmission frame, the BS needs to conduct device activity detection (DAD), channel estimation (CE), and data detection (DD). A possible solution to this problem is to divide each transmission frame into a pilot phase and a data phase. The pilot phase is for pilot transmission at the devices and joint DAD and CE at the BS, and the data phase is for data transmission at the devices and DD at the BS. Since the device activity patterns are sporadic, at any given time, only a small and random fraction of all devices are active. Joint DAD and CE can be cast into a compressed sensing (CS) problem, where advanced compressed sensing algorithms, such as approximate message passing (AMP) \cite{AMP}, sparse Bayesian learning (SBL) \cite{SBL}, turbo compressed sensing (Turbo-CS) \cite{TCS}, and variance state propagation (VSP) \cite{vsp}, can be used to solve the problem. For example, the authors in \cite{ampgf} formulated the joint DAD and CE problem as a compressed sensing  multiple measurement vector (MMV) problem by assuming multiple receive antennas, and used the AMP algorithm to solve the formulated problem. 
% In \cite{AMP-UAD}, the vector AMP algorithm was used to jointly detect active devices and estimate the multiple-input multiple-output (MIMO) channels, where both the transmitters and the receiver are equipped with multiple antennas. 
In \cite{TCS-UAD}, the turbo generalized MMV (GMMV)  algorithm was proposed to solve the joint DAD and CE problem in a MIMO system with mixed analog-to-digital converters.  
In addition to  user sparsity, the channel sparsity in the angle domain of the receiver antenna array can also help with joint DAD and CE. For example, \cite{angle} exploited   user-angle-domain sparsity  in a massive MIMO grant-free system and proposed a Turbo-GMMV-AMP algorithm for the problem. Moreover, machine learning technologies have been applied to further improve the performance.  
% Ref. \cite{amp-asy}  considered an asynchronous system and combined deep learning in the AMP framework to address the joint DAD and CE problem. 
In \cite{dnn}, the authors used a deep neural network to learn the weights involved in message passing to improve the convergence performance. 
In contrast to the two-phase approach, another line of research proposed an one-phase approach \cite{SD1,SD2}, where the BS is required to conduct joint DAD, CE and DD. As compared to the two-phase approach, the more challenging one-phase approach generally increases the computational complexity, but can achieve significant performance improvement by efficiently exploiting the structure (such as sparsity and low rank) inherent in the channels and the signals.

Due to the limited coverage of terrestrial BSs, the development of terrestrial IoT is highly restricted in extreme environments such as deserts, forests, 
and oceans. Recently, satellite is considered as a potential solution for global IoT services \cite{remote-iot, mmtc, survey}. %overview1, 5G, SIN,
In particular, low earth orbit (LEO) satellites supplement and extend terrestrial IoT systems, which can effectively solve the environmental constraints faced by terrestrial networks. In addition, the LEO satellite-IoT system has the ability to be immune to natural disasters and to guarantee all-weather communication. Compared with geostationary orbit (GEO) satellites, LEO satellites operate in low-earth orbits with a height typically lower than 2000 km. A shorter communication distance provides a more real-time IoT service. Yet, the LEO satellite raises new features due to the high speed and the long distance between satellites and ground devices:
\begin{itemize}
    \item Large transmission delay: The delay of LEO satellite channel\footnote{The delay refers to the transmission time of the electromagnetic signals during the space
between the device and the satellite.}, though  much shorter than that of a GEO satellite, is typically more than 1 ms, and is still quite large compared with that in terrestrial  networks. Ref. \cite{dop1} and Ref. \cite{dop2} proposed to use global navigation satellite system (GNSS) based techniques  to synchronize the devices and the satellite, and the delay of each device can be largely compensated. The residual delay can be  handled by existing techniques such as the use of cyclic prefix (CP) in the orthogonal frequency division multiplexing (OFDM) system.
    \item Severe Doppler effect: It has been reported in \cite{dopcal} that in a LEO satellite communication system with a carrier frequency of Ku band, the maximum Doppler shift can be over 200 kHz. 
    In \cite{uad-ce}, the authors considered GF-NORA  in LEO satellite-IoT, and  assumed that with the help of terrestrial BSs,  the Doppler shifts are completely compensated.  But this method is not applicable in remote areas without terrestrial BS. In addition,  with GNSS, the devices can acquire their position information and calculate the Doppler  shifts. However, the compensation of the Doppler shifts at the terrestrial device is incomplete, since there are typically more than one path and a terrestrial device can only compensate the Doppler shift of one path.  The residual Doppler shifts of Ku-band signals can be over several thousand Hertz.  As such, it is of pressing interest to design a grant-free random access scheme that can reliably handle the severe Doppler effect of the LEO satellite-IoT channel.
\end{itemize}

% Thus, it is challenging to guarantee reliable random access of a huge number of  devices to a LEO satellite in such a hostile communication environment.

% Initial attempts to tackle the above problem are  as follows. .  Ref. \cite{Schedule}  used the terrestrial BS to group the devices so that the devices in a group have ignorable differential Doppler shifts.

In this paper, we assume that  with the aid of GNSS, the device delays of the satellite channel can be largely compensated. Then we adopt the OFDM technique  to deal with the residual delays, providing that the residual delays do not exceed the length of  CP. Besides, since the Doppler shift compensation at the IoT devices is expensive and inaccurate,  we propose to deal with the Doppler effect at the satellite by assuming that the average Doppler shift of the devices in a beam is estimated and then compensated.
 We focus on the joint DAD and CE problem in an GF-NORA for LEO satellite-IoT, where active devices suffer from the residual Doppler effect. 
 
 To distinguish the  Doppler components  with high precision, we adopt the OFDM-symbol repetition technique for the pilot design, where a super OFDM symbol are constructed by concatenating repeated regular OFDM symbols. In \cite{zc1, zc2}, the authors designed the long preamble sequence  by concatenating the circularly shifted replicas of a short Zadoff–Chu (ZC) sequence,  for the random access in satellite communication. In addition, similar repetition  techniques have been  used for carrier frequency offset estimation for a single user, while the receiver carries out the  maximum Likelihood (ML)  estimation of the carrier frequency offset  \cite{cof1,cof2,cof3,cof4}. The problem considered in this paper is more challenging. On one hand, due to the multi-path effect, each IoT device generally has more than one Doppler component. On the other hand,  there are a large number of devices in the  satellite-IoT system. As such,   the  ML-based estimation methods, if applied directly,  may incur a prohibitively high computational complexity. 
 
 To estimate the time-varying channel of satellite-IoT, we represent the channel with a grid-based parametric model, and point out that the time-varying OFDM channel exhibits block sparsity in the delay-Doppler domain. Then, together with the sparsity in the user domain (due to sporadic transmission of the terrestrial devices), we formulate the joint DAD and CE problem for OFDM-based GF-NORA in LEO satellite-IoT as a sparse signal recovery problem. %Interestingly, 
  Many existing compressed sensing algorithms \cite{OMP, AMP, GAMP, TCS, STCS, SBL, PCSBL,vsp}  can be applied to provide approximate solutions to the problem. It is known that Bayesian CS algorithms, such as AMP and Turbo-CS, can achieve significant performance improvement over non-Bayesian methods in sparse signal reconstruction. But AMP and Turbo-CS generally rely on a certain randomness property of the measurement matrix to ensure convergence, and the recovery performance of these algorithms may degrade seriously when such randomness is not met.  
  
    As inspired by the robustness of the VSP algorithm  to a broad family of measurement matrices, we extend the VSP algorithm to the massive connectivity scenario by appropriately handling the user sparsity prior, with the resulting algorithm termed modified VSP (MVSP). Specifically, we characterize the channel sparsity structure in the delay-Doppler-user domain with a probability model, which consists of a linear module and a Markov random field (MRF) module. The linear module handles the linear constraint between received signal and unknown vector, and the MRF module  handles the  block-sparse prior in the delay-Doppler domain as well as the sparse prior in the user domain. Different from the original Ising model \cite{ising} in the MRF, we introduce an auxiliary variable  to characterize the relationship between the channel states and the device activity. The two modules are iterated until convergence. 
    The proposed approach generally suffers from the energy leakage problem since the employed parametric channel model is based on a two-dimensional grid in the delay-Doppler domain. To reduce the mismatch between the actual channel and its on-grid representation, an expectation-maximization (EM) based learning method, named EM-MVSP, is proposed to update the delay-Doppler grid parameters.
    % ,  and can be appropriately integrated into the MVSP algorithm, yielding the so-called EM-MVSP algorithm. 
 The contributions of this paper are summarised as follows.

\begin{itemize}
    
    \item  We develop a grid-based parametric system model for the OFDM-based satellite-IoT, and formulate the joint DAD and CE problem  as a sparse signal recovery problem. Interestingly, we show that the measurement matrix in our considered problem can only be partially manipulated by the design of pilots, and exhibits a special correlation structure caused by the Doppler effect.  Our experiments show that most existing compressed sensing algorithms including AMP and Turbo-CS behave poorly in the considered problem.

    \item To distinguish the  Doppler components, we adopt  the OFDM-symbol repetition technique  to increase the frequency resolution of the OFDM system. We  show that this OFDM-symbol repetition technique can efficiently improve the DAD and CE performance of the OFDM-based satellite-IoT system.
    
    \item We propose the MVSP algorithm for the joint DAD and CE problem, which is robust to the measurement matrix in our problem.  Different from the original VSP algorithm, we introduce an auxiliary variable  to characterize the relationship between the channel states and the device activity, and thus the DAD can be conducted by the MRF module. 
    
    \item To alleviate the mismatch of the grid-base model, we further propose the EM-MVSP algorithm to update the grid parameters using EM method.
    We show that significant performance improvement can be achieved by the EM-MVSP algorithm, as compared to the counterpart algorithms including AMP, SBL, Turbo-CS and MVSP.
 
\end{itemize}

The rest of the paper is organized as follows. Section II introduces the time-varying satellite- IoT channel and the GF-NORA satellite-IoT system model, and then transforms it to a parametric form. Section III formulates the DAD and CE problem, constructs a probability model, and presents the MVSP algorithm. In Section IV, the MVSP algorithm is extended to the mismatch scenario with the EM framework. Numerical results are given in Section V, and Section VI concludes this paper. The frequently used abbreviations  in this paper are summarized in Table \ref{abbr}.

\textit{Notation:} We use a bold symbol lowercase letter and bold symbol capital letter to denote a vector and a matrix, respectively. 
The trace, transpose, conjugate transpose, inverse, vectorization and Frobenius norm of a matrix are denoted by $\text{Tr}(\cdot)$, $(\cdot)^{\text{T}}$, $(\cdot)^{\text{H}}$, $(\cdot)^{-1}$, $\text{vec}(\cdot)$ and $\|\cdot\|_{\text{F}}$, respectively; 
$\propto$ represents both sides of the equation are multiplicatively connected to a constant; $\text{diag}(\boldsymbol{a})$ forms a diagonal matrix with the diagonal elements in $\boldsymbol{a}$;
$|\cdot|$ denotes the modulus of a complex number;  $\|\cdot\|$ denotes the $\ell^2$ norm;  $\mathbb{E}_p[\boldsymbol{x}]$ denotes the expectation of $\boldsymbol{x}$ with respect to distribution $p$;
$\otimes$ denotes the Kronecker product;
$[N]$ denotes the set $\{1,2,\ldots, N\}$; 
$\delta(\cdot)$ denotes the Dirac delta function; 
%  $\mathcal{O}$ is the big-O notation.
The Gaussian  and complex Gaussian distribution of $\boldsymbol{x}$ with mean
$\bar{\boldsymbol{m}}$ and covariance matrix $\boldsymbol{\Sigma}$ is denoted by ${\cal{N}}(\boldsymbol{x}; \bar{\boldsymbol{m}},\boldsymbol{\Sigma})$ and  ${\cal{CN}}(\boldsymbol{x}; \bar{\boldsymbol{m}},\boldsymbol{\Sigma})$, respectively.

\begin{table}
\center
\caption{Frequently used abbreviations and corresponding meaning.}
\begin{tabular}{ |m{5em}| m{5cm}| m{5em}| m{5cm} | }
\hline
 Abbr. & Meaning & Abbr. & Meaning \\ 
\hline
 AMP & Approximate message passing & CE  & Channel estimation \\
\hline 
CP  &  Cyclic prefix &  CS & Compressed sensing \\
\hline
DAD & Device activity detection &  EM &  Expectation–maximization\\
\hline
 GAMP & Generalized approximate message passing  & GF-NORA & Grant-free non-orthogonal random access\\
\hline
GNSS & Global navigation satellite system & IoT & Internet of Things \\  
\hline
LEO & Low earth orbit &  ML &  Maximum Likelihood\\
\hline 
MRF  & Markov random field & MVSP & Modified variance state propagation    \\
\hline
NMSE & Normalized mean-squared error & OFDM & Orthogonal frequency division multiplexing   \\
\hline
OMP  & Orthogonal matching pursuit & PCSBL  & Pattern-coupled sparse Bayesian learning  \\
\hline
SBL & Sparse Bayesian learning & SNR &  Signal-noise-ratio  \\
\hline
STCS & Structured turbo compressed sensing & VSP & Variance state propagation  \\
\hline
\end{tabular}
\label{abbr}
\end{table}

\section{System Model}
\subsection{Time-Varying Satellite-IoT Channel}
    \begin{figure}[htbp]
        \centering
        \includegraphics[width=2.4in]{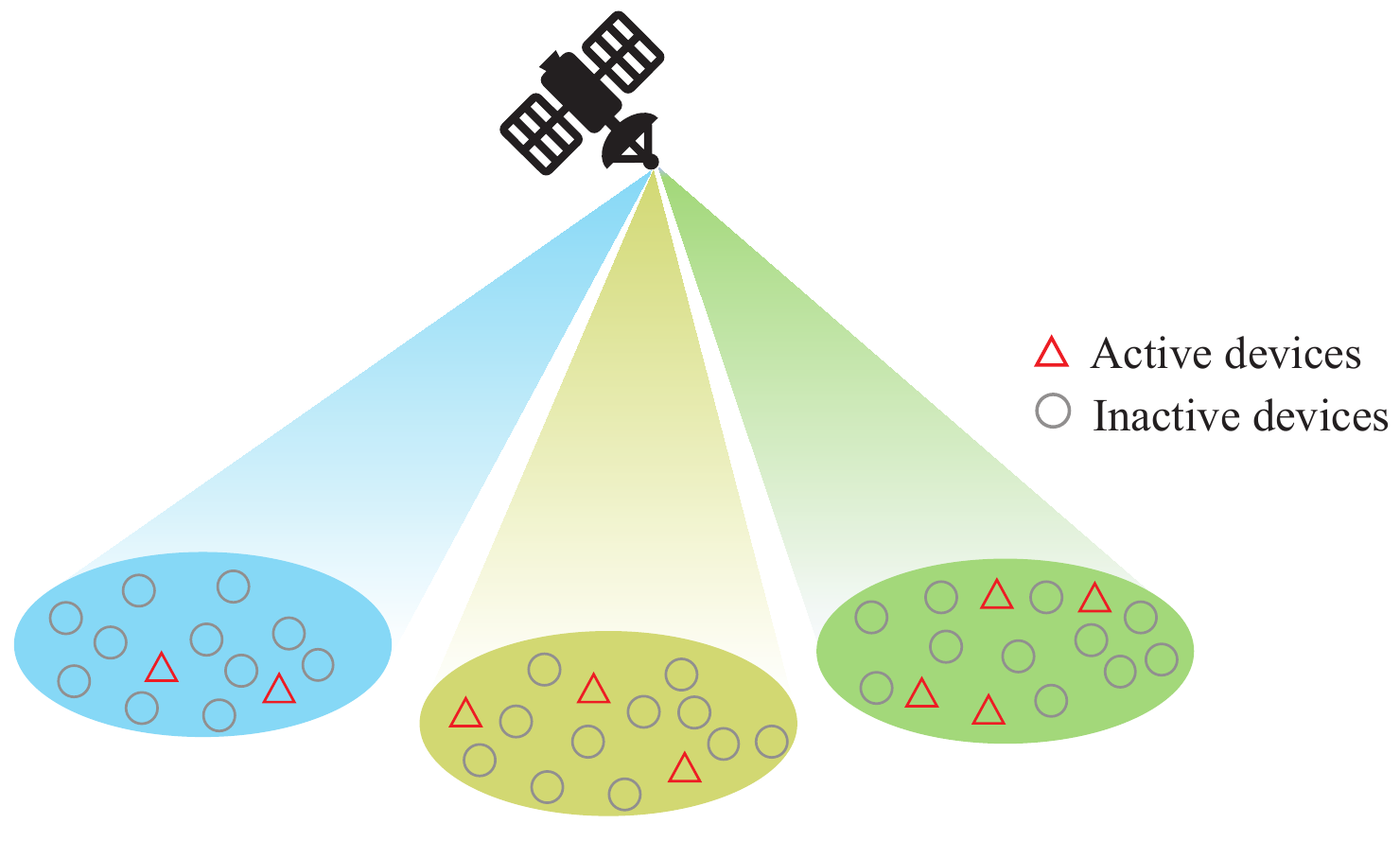}
        \caption{The LEO satellite-IoT model.}
        \label{system-model}
    \end{figure}
    As illustrated in Fig.\ \ref{system-model}, we consider a multi-beam LEO-satellite  system where there exists $K$ potential devices within a beam coverage. 
    We assume that the signals in different beams are orthogonal, i.e.,  the inter-beam interference is ignored. Then, within a beam, the noiseless baseband received scalar signal from device $k$ at the satellite can be expressed as
    \begin{equation}\label{rk1}
        r_k(t) = \int_{-\infty}^{\infty} \int_{-\infty}^{\infty} \bar{h}_k(\tau, \nu)s_k(t-\tau) e^{j2\pi\nu t} {\rm d}\tau {\rm d}\nu ,
    \end{equation}
    where $\bar{h}_k(\tau,\nu)$ is the channel impulse response at delay $\tau$ and Doppler frequency $\nu$, and $s_k(t)$ is the signal from the $k$-th device. As different devices are usually geographically separated, we  assume that the channel realizations between the satellite and different devices are uncorrelated. We also assume that with the help of GNSS, the delay can be partially pre-compensated at the device, and the satellite uses the Doppler frequency shift at the beam center to compensate all the devices in the beam. As such,  the impact of the transmission delay and Doppler frequency shift can be largely eliminated, or in other words, only the residual delay and the residual Doppler frequency shift need to be taken into account.  Without loss of generality, we assume that there exist $P_k$ paths for each device $k$, where $P_k$ is a small positive integer since the satellite communication  is typically in a weak multipath transmission environment. The residual delay and the residual Doppler shift of the $p$-th path of device $k$ are denoted as $\Bar{\tau}_{k,p}$ and $\Bar{\nu}_{k,p}$, respectively. We assume that $\Bar{\tau}_{k,p}$ and $\Bar{\nu}_{k,p}$ are constant during a transmission frame. The corresponding attenuation including path loss, reflection and processing gains of each device in the $p$-th path is characterized by a complex coefficient $\bar{h}_{k,p}$. Therefore, $\bar{h}_k(\tau,\nu)$ can be approximated by
    \begin{equation}\label{h-spar}
        \bar{h}_k(\tau,\nu) = \sum\limits_{p=0}^{P_k-1} \bar{h}_{k,p} \delta(\tau- \Bar{\tau}_{k,p}) \delta(\nu- \Bar{\nu}_{k,p}).
    \end{equation}
    Substituting \eqref{h-spar} into \eqref{rk1} yields
    \begin{equation}\label{rk2}
        r_k(t) = \sum\limits_{p=0}^{P_k-1} \bar{h}_{k,p} s_k(t-\Bar{\tau}_{k,p}) e^{j2\pi\Bar{\nu}_{k,p} t}.
    \end{equation}
    \subsection{Grant-Free Satellite-IoT System Model}

    In this paper, we adopt the grant-free non-orthogonal random access (GF-NORA) scheme, in which the devices share the physical channel resource and directly transmit their signals without requiring the permission of the satellite.
    Following  \cite{GF}, we assume that each transmission frame in GF-NORA consists of   two phases, namely,  the pilot phase and the data phase. Each device is preassigned with a unique non-orthogonal pilot sequence. In the pilot phase, the active devices transmit their pilots, based on which  the satellite detects the active devices and estimates their channels. In the data phase, the devices transmit data  without the grant from the satellite, and the  satellite decodes the data based on  the estimated channel of the active devices. Orthogonal frequency division multiplexing (OFDM) is employed for both the pilot and data transmission phases. 
    In this work, we  focus on the device activity detection (DAD) and channel estimation (CE) in the pilot phase. The system model of the pilot phase is described as follows.
 
\begin{figure*}[htbp]
        \centering
        \includegraphics[width=4.5in]{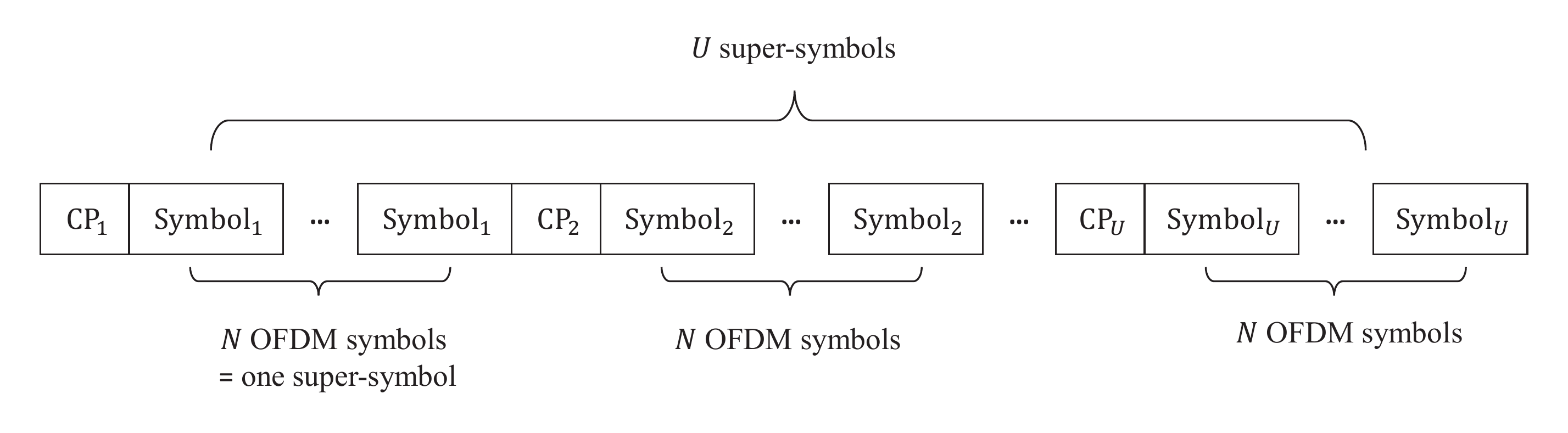}
        \caption{The structure of a pilot sequence in the pilot phase.}
        \label{frame}
    \end{figure*}

 In the pilot phase, we construct a super-symbol by concatenating $N$ repetitions of an OFDM symbol, and a cyclic prefix (CP) is  applied to eliminate the inter-symbol interference, as shown in Fig.~\ref{frame}. The duration of such a super-symbol with a CP is $\Bar{T} = NT+ T_{\text{cp}}$, where $T$ is the length of a regular OFDM symbol, $T_{\text{cp}}>\tau_{\text{max}}$ is the length of the CP, and $\tau_{\text{max}}$ is the maximal residual delay for all devices. We note that the above repetitions of an OFDM symbol improve the frequency resolution of the receiver, so that the Doppler frequency shifts can be identified more clearly and thus its adverse effect can be more efficiently mitigated. Without loss of generality, we assume that the pilot sequence of a device  contains $U$ consecutive super-symbols. The frequency spacing between any two adjacent subcarriers is set to $\Delta f = 1/T$. In the $u$-th super-symbol, the baseband modulated signal at the $k$-th device is given by
 \begin{subequations}
     \begin{equation}\label{d}
        d_{k,u}(t) = \sum\limits_{m=0}^{M-1} x_{k,m,u} e^{j2\pi m \Delta f t} \xi(t-u\Bar{T}), \forall k, \forall u,
    \end{equation}
    where $x_{k,m,u}$ is the pilot on the $m$-th subcarrier in the $u$-th  super-symbol of the $k$-th device, and 
    \begin{equation}\label{win}
        \xi(t)=\begin{cases}
    1, &t\in \left[-T_{\text{cp}}, NT\right],\\
    0, &\text{otherwise},
    \end{cases}
    \end{equation}
    is the transmitted rectangular  pulse. The baseband pilot signal of device $k$ 
    is given by
    \begin{equation}\label{s}
        s_k(t) = \sum\limits_{u=0}^{U-1}d_{k,u}(t), \forall k,
    \end{equation}
    where $U$ is the number of super-symbols in a pilot sequence. 
 \end{subequations}
    
    In each transmission frame, only a small subset of devices are active. To characterize such sporadic transmission, the device activity is represented by an indicator function $\alpha_k$ as
    \begin{equation}\label{alpha}
        \alpha_k=\begin{cases}
    1, &\text{if device $k$ is active},\\
    0, &\text{if device $k$ is inactive},
    \end{cases}\quad k = 1,\cdots, K,
    \end{equation}
 with a probability $p(\alpha_k=1) = \rho $ where $\rho \ll 1$.
  Combining  \eqref{rk2}, \eqref{d}, \eqref{s} and \eqref{alpha}, 
% %   the  received baseband signal of device $k$ at the satellite is given by
%      \begin{align}\label{eq.8}
%         r_k(t) = &\alpha_k\sum\limits_{p=0}^{P_k-1} e^{j2\pi\Bar{\nu}_{k,p} t} \bar{h}_{k,p}\sum\limits_{u=0}^{U-1} \sum\limits_{m=0}^{M-1}x_{k,m,u}    e^{j2\pi m \Delta f (t-\Bar{\tau}_{k,p})} \xi(t-\Bar{\tau}_{k,p}-u\Bar{T}).
%     \end{align}
%     Hence,  
    the  received baseband signal of all devices  at the satellite can be expressed as
    \begin{align}\label{eq.9}
        r(t) =& \sum\limits_{k=1}^K r_k(t) +w(t) \nonumber\\
        =& \sum\limits_{k=1}^K\alpha_k \sum\limits_{p=0}^{P_k-1} e^{j2\pi\Bar{\nu}_{k,p} t} \bar{h}_{k,p} \sum\limits_{u=0}^{U-1}\sum\limits_{m=0}^{M-1} x_{k,m,u} e^{j2\pi m \Delta f (t-\Bar{\tau}_{k,p})} \xi(t-\Bar{\tau}_{k,p}-u\Bar{T})  + w(t),
    \end{align}
    where $w(t)$ is an additive white Gaussian noise (AWGN).
    
    In a conventional OFDM system with  symbol duration $T$, any two subcarriers with minimum frequency shift $\Delta f = 1/T$ are orthogonal to each other, i.e., the frequency resolution is the subcarrier spacing $1/T$. In our super-symbol system, since each super-symbol consists of  $N$ repeated  regular OFDM symbols, the frequency resolution is  $1/(NT)$. In other words, $N$-time oversampling in the frequency domain can be  applied in our system.
    In the $u$-th  super-symbol interval, the   demodulated signals  with $N$-time oversampling is given by
    \begin{align}\label{y1}
        y_{n,u} = \frac{1}{NT}\int_{u\Bar{T}}^{u\Bar{T}+NT} &r(t)e^{-j2\pi\frac{n}{N}\Delta f t}{\rm d}t, \;\;\text{for} \;\;n \in\{0,1,\ldots,NM-1\},
    \end{align}
    where  the boundary of the observation window of the $u$-th super-symbol  specifies  the upper and  lower limits of the integral. By plugging \eqref{eq.9} into \eqref{y1}, $y_{n,u}$ can be expressed as
    \begin{subequations}
    \begin{align}\label{y2}
        y_{n,u} =&  \sum\limits_{k=1}^K\frac{\alpha_k}{NT}\sum\limits_{m=0}^{M-1} x_{k,m,u}\sum\limits_{p=0}^{P_k-1}e^{-j2\pi m \Delta f \Bar{\tau}_{k,p}}\bar{h}_{k,p}\int_{u\Bar{T}}^{u\Bar{T}+NT}e^{j2\pi\Bar{\nu}_{k,p} t} e^{j2\pi  (m-\frac{n}{N}) \Delta f t} {\rm d}t+w_{n,u} \nonumber\\
        =&  \sum\limits_{m=0}^{M-1}\sum\limits_{k=1}^K x_{k,m,u}g_{m,n,k,u}+w_{n,u},
    \end{align}
    where $w_{n,u} = \int_{u\Bar{T}}^{u\Bar{T}+NT} e^{j2\pi \frac{n}{N}\Delta f t}w(t){\rm d}t$, and 
    \begin{align} \label{g}
        g_{m,n,k,u} = & \frac{\alpha_k}{NT}\sum\limits_{p=0}^{P_k-1}e^{-j2\pi m \Delta f \Bar{\tau}_{k,p}}\bar{h}_{k,p}  \int_{u\Bar{T}}^{u\Bar{T}+NT}e^{j2\pi\Bar{\nu}_{k,p} t} e^{j2\pi(m-\frac{n}{N})\Delta f t} {\rm d}t\text{.}
    \end{align}
    \end{subequations}
    Let $\boldsymbol{x}_{m,u} = \left[x_{1,m,u},  \cdots, x_{K,m,u}\right]^T$ and $\boldsymbol{g}_{m,n,u}=\left[g_{m,n,1,u},  \cdots, g_{m,n,K,u}\right]^T$. We can rewrite $y_{n,u}$ %in \eqref{y2} 
    as
    \begin{equation}\label{y3}
        y_{n,u} = \sum\limits_{m=0}^{M-1}\boldsymbol{x}_{m,u}^T \boldsymbol{g}_{m,n,u} + w_{n,u}.
    \end{equation}
    Let $\boldsymbol{y}_u=[y_{0,u},\cdots, y_{N\!M-1,u}]^T$. We have
    \begin{subequations}
    \begin{align}\label{y-real}
        \boldsymbol{y}_u
        %&=\sum\limits_{m=0}^{M-1}\begin{bmatrix} \boldsymbol{x}_{m,u}^T \boldsymbol{g}_{m,0,u}\\ \vdots \\ \boldsymbol{x}_{m,u}^T \boldsymbol{g}_{m,N\!M-1,u} \\ \end{bmatrix} + \begin{bmatrix} w_{0,u}\\ \vdots \\ w_{NM-1,u} \\ \end{bmatrix}\nonumber\\
        &=\boldsymbol{G}_u\boldsymbol{x}_u+ \boldsymbol{w}_u,
    \end{align}
    where $\boldsymbol{w}_u = \left[ w_{0,u},\ldots,  w_{NM-1,u}  \right]^{\text{T}}$ is the AWGN with variance $\sigma^2$, $ \boldsymbol{x}_u \triangleq{ \left[\boldsymbol{x}_{0,u}^T,\cdots, \boldsymbol{x}_{M-1,u}^T\right]^T} \!\in \!\mathbb{C}^{MK\times 1}$ ,
    and
    \begin{align}\label{Gu}
        \boldsymbol{G}_u &\triangleq{
        \begin{bmatrix} 
        \boldsymbol{g}_{0,0,u}^T &\cdots &\boldsymbol{g}_{M-1,0,u}^T\\ 
        &\vdots& \\ 
        \boldsymbol{g}_{0,N\!M-1,u}^T &\cdots &\boldsymbol{g}_{M-1,N\!M-1,u}^T \end{bmatrix}
        } \! \in \!\mathbb{C}^{NM\times MK}.
    \end{align}
    \end{subequations}
    $\boldsymbol{G}_u$ characterizes the  channels of all the $M$ subcarriers in the $u$-th super-symbol. Our goal is to estimate $\{\boldsymbol{G}_u\}_{u=0}^{U-1}$ based on  $\{\boldsymbol{y}_u\}_{u=0}^{U-1}$ and  $\{\boldsymbol{x}_u\}_{u=0}^{U-1}$. A brute-force approach to this problem is infeasible since  the number of observations in $\{\boldsymbol{y}_u\}_{u=0}^{U-1}$ is only $UMK$,  which is far less than that of the unknown variables in $\{\boldsymbol{G}_u\}_{u=0}^{U-1}$, i.e.,  $UNM^2K$. As such, we need a more elegant representation of $\{\boldsymbol{G}_u\}_{u=0}^{U-1}$ with less unknowns, as detailed in the next subsection.

    \subsection{Grid-Based Parametric System Model} \label{Paramodel}
    We now present a grid-based parametric model for the channel $\boldsymbol{G}_u$. From \eqref{g} and \eqref{Gu}, the unknown parameters in $\boldsymbol{G}_u$ include each device's  path delays $\{\bar{\tau}_{k,p}\}_{p=0}^{P_k-1}$, path Doppler shifts $\{\bar{\nu}_{k,p}\}_{p=0}^{P_k-1}$ and  channel gains $\{\bar{h}_{k,p}\}_{p=0}^{P_k-1}$.
    In practical wireless communication scenarios, each channel path may consist of many sub-paths, and the parameters of all these sub-paths are usually difficult to distinguish. Exact identification of these parameters is therefore very challenging. 

    To address this issue, we discretize the delay domain and the Doppler domain into a two-dimensional grid. Instead of estimating the equivalent channel matrix $\boldsymbol{G}_u$, we estimate the representation of the channel on the  grid. Then $\boldsymbol{G}_u$ can be recovered based on the grid parameters and the corresponding channel representation. This approach avoids the estimation of each physical path or sub-path separately, but considers the overall representation  of the physical channel on the delay-Doppler grid. In specific, for each device $k$, the grid parameters of the delay domain and the Doppler domain are defined  as
    \begin{align}
        \boldsymbol{\tau}_k &=\left\{\tau_{l,k}\right\}_{l=0}^{L-1},~\tau_{l,k}\in\left[0, \tau_{\text{max}}\right), \nonumber\\
        \boldsymbol{\nu}_k &=\left\{\nu_{j,k}\right\}_{j=0}^{J-1},~\nu_{j,k}\in\left[-\nu_{\text{max}}/2, \nu_{\text{max}}/2\right).\nonumber
    \end{align}
    %We assume that different devices have the same sampled grid. 
    As such, the  channel $\bar{h}_k(\tau,\nu)$ can be approximated as
    \begin{equation}\label{h-grid}
        \bar{h}_k(\tau,\nu) = \sum\limits_{l=0}^{L-1} \sum\limits_{j=0}^{J-1} h_{k,l,j} \delta(\tau- \tau_{k,l}) \delta(\nu- \nu_{k,j}),
    \end{equation}
    where $h_{k,l,j}$ is device $k$'s  channel representation at grid $(\tau_{k,l}, \nu_{k,j})$.
    We rewrite $y_{n,u}$ in \eqref{y2}  as 
     \begin{align}\label{eq.17}
        y_{n,u} \!=& \frac{1}{NT}\int_{u\Bar{T}}^{u\Bar{T}+NT} r(t)e^{-j2\pi \frac{n}{N}\Delta f t}{\rm d}t\nonumber\\
         \!= \!&   \sum\limits_{k=1}^K\frac{\alpha_k}{NT}\sum\limits_{m=0}^{M-1} x_{k,m,u} \sum\limits_{j=0}^{J-1} \! \int_{u\Bar{T}}^{u\Bar{T}+NT} \!\!\! e^{j2\pi\nu_{k,j} t}  e^{j2\pi (m-\frac{n}{N}) \Delta f t} {\rm d}t \sum\limits_{l=0}^{L-1} e^{-j2\pi m \Delta f \tau_{k,l}}h_{k,l,j} + w_{n,u}.
    \end{align}
    Let $\boldsymbol{b}_{k,m} = \left[e^{-j2\pi m \Delta f \tau_{k,0}}, \cdots, e^{-j2\pi m \Delta f \tau_{k,L-1}}\right]^{\text{T}}\in \mathbb{C}^{L\times 1}$, $\boldsymbol{h}_{k,j} = \left[h_{k,0,j}, \cdots, h_{k,L-1,j}\right]^T\in \mathbb{C}^{L\times 1}$, and $c_{k,m,n,j,u}=\frac{1}{NT}\int_{u\Bar{T}}^{u\Bar{T}+NT}e^{j2\pi\nu_{k,j} t} e^{j2\pi (m-\frac{n}{N}) \Delta f t} {\rm d}t$. We rewrite \eqref{eq.17} as
    \begin{subequations}
    \begin{align}\label{eq.18}
        y_{n,u} =& \sum\limits_{k=1}^K\alpha_k\sum\limits_{m=0}^{M-1} x_{k,m,u} \sum\limits_{j=0}^{J-1} c_{k,m,n,j,u} \boldsymbol{b}_{k,m}^T\boldsymbol{h}_{k,j}+w_{n,u}  \nonumber\\
        % =& \sum\limits_{k=1}^K\alpha_k\sum\limits_{m=0}^{M-1} x_{k,m,u} \boldsymbol{b}_{k,m}^T \boldsymbol{H}_k \boldsymbol{c}_{k,m,n,u}+w_{n,u} \nonumber\\
        =& \sum\limits_{k=1}^K\alpha_k \left[\sum\limits_{m=0}^{M-1} x_{k,m,u} \left(\boldsymbol{b}_{k,m}^T \otimes \boldsymbol{c}_{k,m,n,u}^T\right)\right]  \textup{vec}\left(\boldsymbol{H}_k^T\right)+w_{n,u} \nonumber\\
        =& \sum\limits_{k=1}^K\alpha_k \boldsymbol{a}_{k,n,u} \textup{vec}\left(\boldsymbol{H}_k^T \right)+w_{n,u},
    \end{align}
    where $\boldsymbol{c}_{k,m,n,u} = \left[c_{k,m,n,0,u}, \cdots, c_{k,m,n,J-1,u}\right]^T\in \mathbb{C}^{J\times 1}$, $\boldsymbol{H}_k = \left[\boldsymbol{h}_{k,0}, \cdots, \boldsymbol{h}_{k,J-1}\right]\in \mathbb{C}^{L\times J}$, and 
    \begin{align} \label{ak}
        \boldsymbol{a}_{k,n,u} = \sum\limits_{m=0}^{M-1} x_{k,m,u} \left(\boldsymbol{b}_{k,m}^T \otimes \boldsymbol{c}_{k,m,n,u}^T\right)\in \mathbb{C}^{1\times LJ}.
    \end{align}
    \end{subequations}
    Then we have
    \begin{subequations}
    \begin{align}\label{yu}
        \boldsymbol{y}_u
        %&=\sum\limits_{k=1}^{K}\begin{bmatrix} \alpha_k \textup{vec}\left(\boldsymbol{H}_k^T\right)^T\boldsymbol{a}_{k,0,u}^T\\ \vdots \\ \alpha_k \textup{vec}\left(\boldsymbol{H}_k^T\right)^T\boldsymbol{a}_{k,N\!M-1,u}^T \\ \end{bmatrix} + \begin{bmatrix} w_{0,u}\\ \vdots \\ w_{N\!M-1,u} \\ \end{bmatrix}\nonumber\\
        &=\boldsymbol{A}_u\boldsymbol{h}+ \boldsymbol{w}_u,
    \end{align}
    where  $\boldsymbol{h}_u \triangleq{ \left[\alpha_1 \textup{vec}\left(\boldsymbol{H}_1^T\right)^T,\cdots, \alpha_K \textup{vec}\left(\boldsymbol{H}_K^T\right)^T\right]^T} \in \mathbb{C}^{Q\times 1}$, 
    \begin{align}
        \boldsymbol{A}_u &\triangleq{\begin{bmatrix} 
        \boldsymbol{a}_{1,0,u} & \cdots & \boldsymbol{a}_{K,0,u}\\ 
        & \vdots & \\ 
        \boldsymbol{a}_{1,N\!M-1,u} & \cdots & \boldsymbol{a}_{K,N\!M-1,u} \\ \end{bmatrix}}\in \mathbb{C}^{NM\times Q} , \label{Au}
    \end{align}
    \end{subequations}
     $R=NMU$ and $Q=KLJ$. $\boldsymbol{h}$ combines the device activity and  delay-Doppler grid parameters. Note that $\boldsymbol{h}$ is a sparse vector since the channel is sparse in the user-delay-Doppler domain, as elaborated in Section \ref{spar}.
    
    Denote by $\boldsymbol{y} \triangleq{\left[\boldsymbol{y}_0^T, \cdots, \boldsymbol{y}_{U-1}^T\right]^T}\in \mathbb{C}^{R\times 1}$ the collection of  all the $U$ demodulated super-symbols in the pilot phase. From \eqref{yu}, we obtain
    \begin{align}\label{y-grid}
        \boldsymbol{y} = \boldsymbol{A} \boldsymbol{h} + \boldsymbol{w},
    \end{align}
    where $\boldsymbol{A} = \left[\boldsymbol{A}_0^T , \cdots, \boldsymbol{A}_{U-1}^T \right]^T\in \mathbb{C}^{R\times Q}$, and $\boldsymbol{w} = \left[\boldsymbol{w}_0^T , \cdots, \boldsymbol{w}_{U-1}^T \right]^T\in \mathbb{C}^{R\times 1}$.
    
    We emphasize that $\boldsymbol{h}$ contains  all the unknown channel parameters  for the reconstruction of the channels $\{\boldsymbol{G}_u\}_{u=0}^{U-1}$. To be specific, we use model \eqref{y-real} to generate channels and signals in simulation, and use the discretized model \eqref{y-grid} to design DAD and CE algorithms.
    We then reconstruct the  channel $\boldsymbol{G}_u$  by the estimated $\boldsymbol{h}$. Suppose that $\Hat{\boldsymbol{h}}$ is an estimate of $\boldsymbol{h}$, $\Hat{h}_{k,l,j}$ is the estimate corresponding to the $(k,l,j)$ grid point, and $\Hat{\alpha}_k$ is an estimate of $\alpha_k$. Then the recovered channel  is given by
    \begin{subequations}
    \begin{equation} \label{G-hat}
        \Hat{\boldsymbol{G}}_u \triangleq{\begin{bmatrix} \Hat{\boldsymbol{g}}_{0,0,u}^T & \cdots & \Hat{\boldsymbol{g}}_{M-1,0,u}^T \\ 
        & \vdots & \\ 
        \Hat{\boldsymbol{g}}_{0,N\!M-1,u}^T & \cdots & \Hat{\boldsymbol{g}}_{M-1,N\!M-1,u}^T  \end{bmatrix}},
    \end{equation}
    where $\Hat{\boldsymbol{g}}_{m,n,u}=\left[\Hat{g}_{m,n,1,u}, \Hat{g}_{m,n,2,u}, \cdots, \Hat{g}_{m,n,K,u}\right]^T$, and
    \begin{align}
        \Hat{g}_{m,n,k,u} =&  \frac{\Hat{\alpha}_k}{NT}\sum\limits_{l=0}^{L-1} \sum\limits_{j=0}^{J-1}e^{-j2\pi m \Delta f \tau_{k,l}}\Hat{h}_{k,l,j} \int_{u\Bar{T}}^{u\Bar{T}+NT}e^{j2\pi\nu_{k,j} t} e^{j2\pi (m-\frac{n}{N}) \Delta f t} {\rm d}t.
    \end{align}
    \end{subequations}
    
    The normalized mean-squared error (NMSE) of the channel estimation is defined by 
    \begin{align} \label{nmse}
        \text{NMSE} =  \frac{1}{U}\sum_{u=0}^{U-1}\frac{\mathbb{E}\left[\|\hat{\boldsymbol{G}}_u-\boldsymbol{G}_u\|_\text{F}^2\right]}{\mathbb{E}\left[\|\boldsymbol{G}_u\|_{\text{F}}^2\right]}.
    \end{align}
    \subsection{Channel Sparsity} \label{spar}
    It is well known that channel sparsity can be exploited to significantly reduce the number of pilots required in channel estimation. In this regard, the channel in our considered satellite-IoT scenario exhibit the following sparsity structure:
    \begin{enumerate}
         \item Sparsity in the user domain:  Most of the devices are inactive at any given time, i.e., most of $\{\alpha_k\}$ are zero. 
        \item Sparsity in the delay-Doppler domain: Since the  satellite  communication   is in a  weak  multi-path transmission environment, the numbers of dominant paths between the satellite and the devices are limited.
        \item Block-sparsity in the delay-Doppler domain: The scattering effect of the electromagnetic waves causes  delay and Doppler  spread in wireless  channels. Besides, the grid mismatch causes additional spreading in the  delay-Doppler domain. These effects make the channel coefficients appear in clusters in the delay-Doppler domain.
    \end{enumerate}
       
\begin{figure}[htbp]
        \centering
        \includegraphics[width=2in]{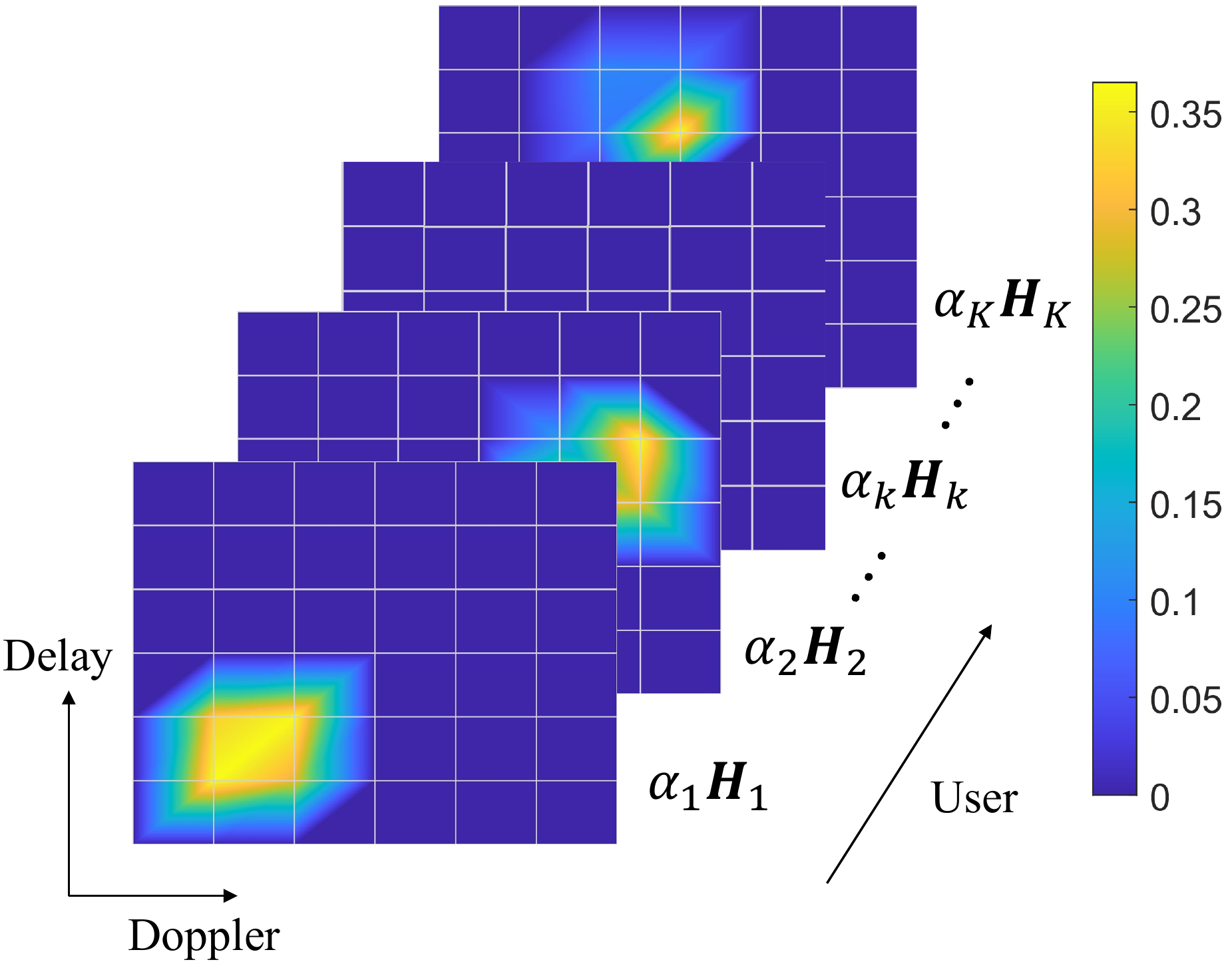}
        \caption{An illustration of the channel sparsity structure.}
        \label{bs}
    \end{figure}

    Fig. \ref{bs} illustrates the sparsity structure of $\boldsymbol{h}$ in the delay-Doppler-user domain. This sparsity structure  is exploited in the  design  of the receiver.
    \subsection{Problem Description} \label{PD}
\begin{figure*}[htbp]
    \centering
    \subfigure[]{\label{ther1}
    \begin{minipage}[t]{0.3\linewidth}
    \centering
    \includegraphics[width=2.2in]{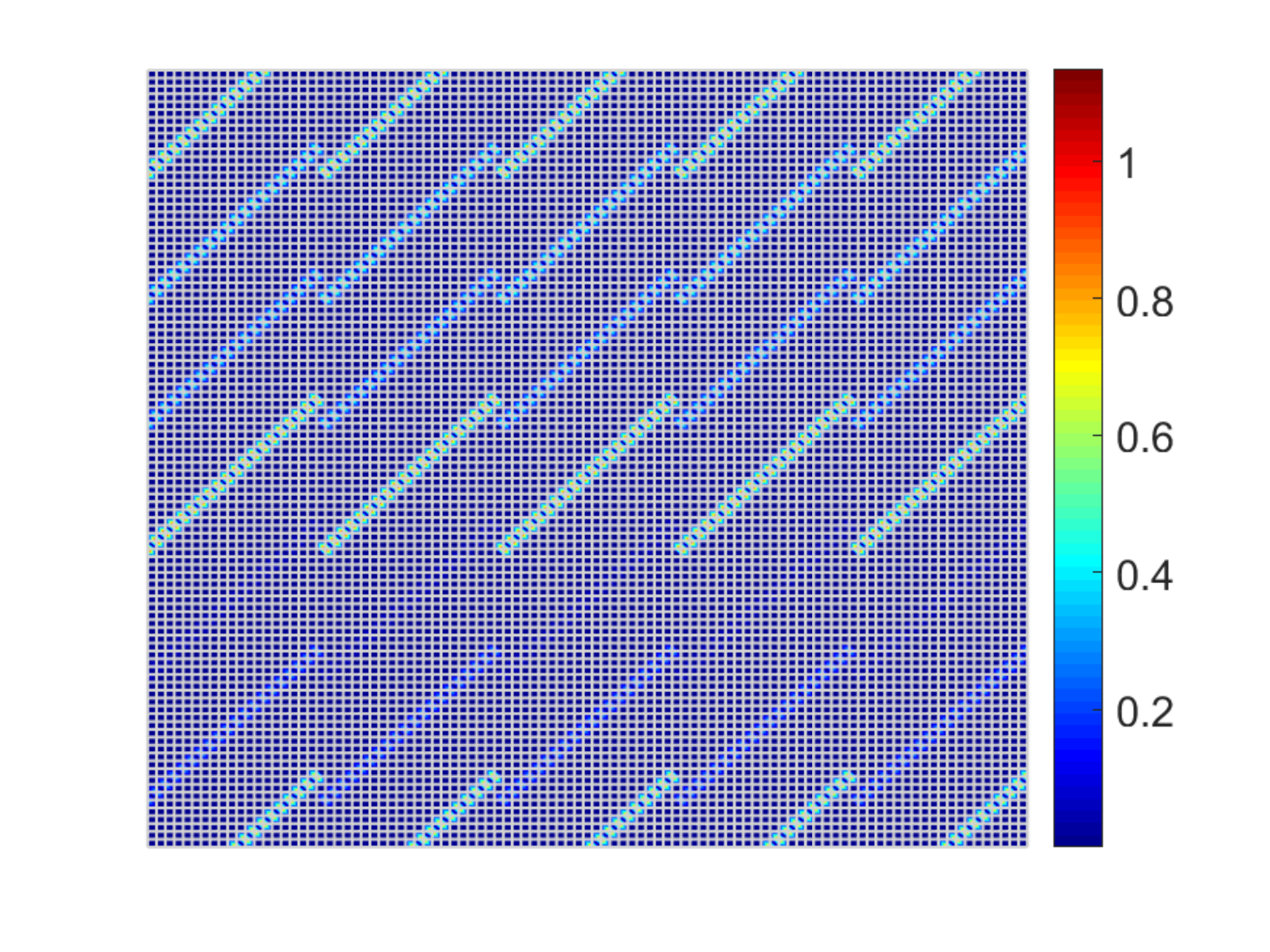}
    \end{minipage}
    }
    \subfigure[]{\label{ther2}
    \begin{minipage}[t]{0.3\linewidth}
    \centering
    \includegraphics[width=2.2in]{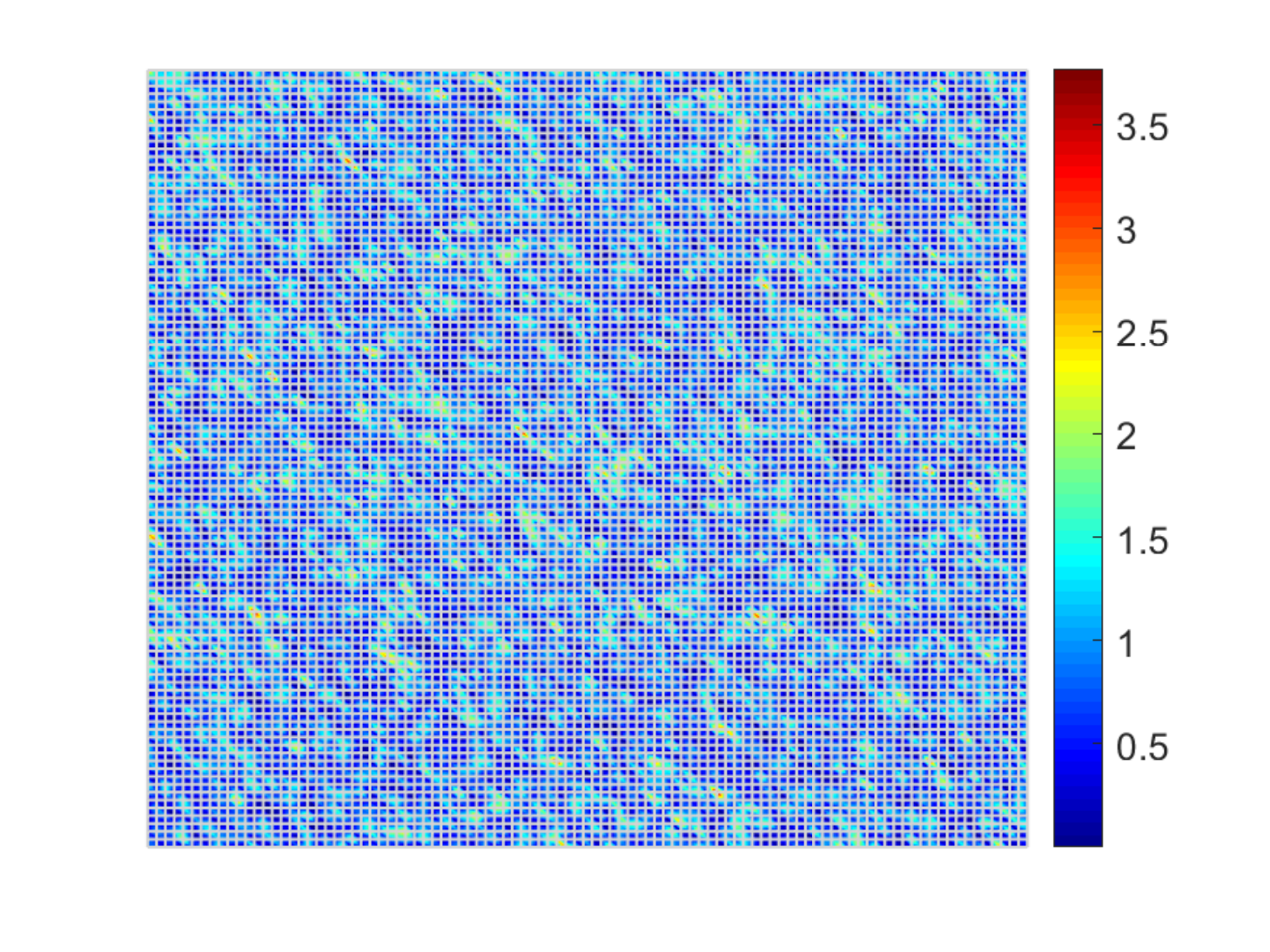}
    \end{minipage}
    }
    \subfigure[]{\label{ther3}
    \begin{minipage}[t]{0.3\linewidth}
    \centering
    \includegraphics[width=2.2in]{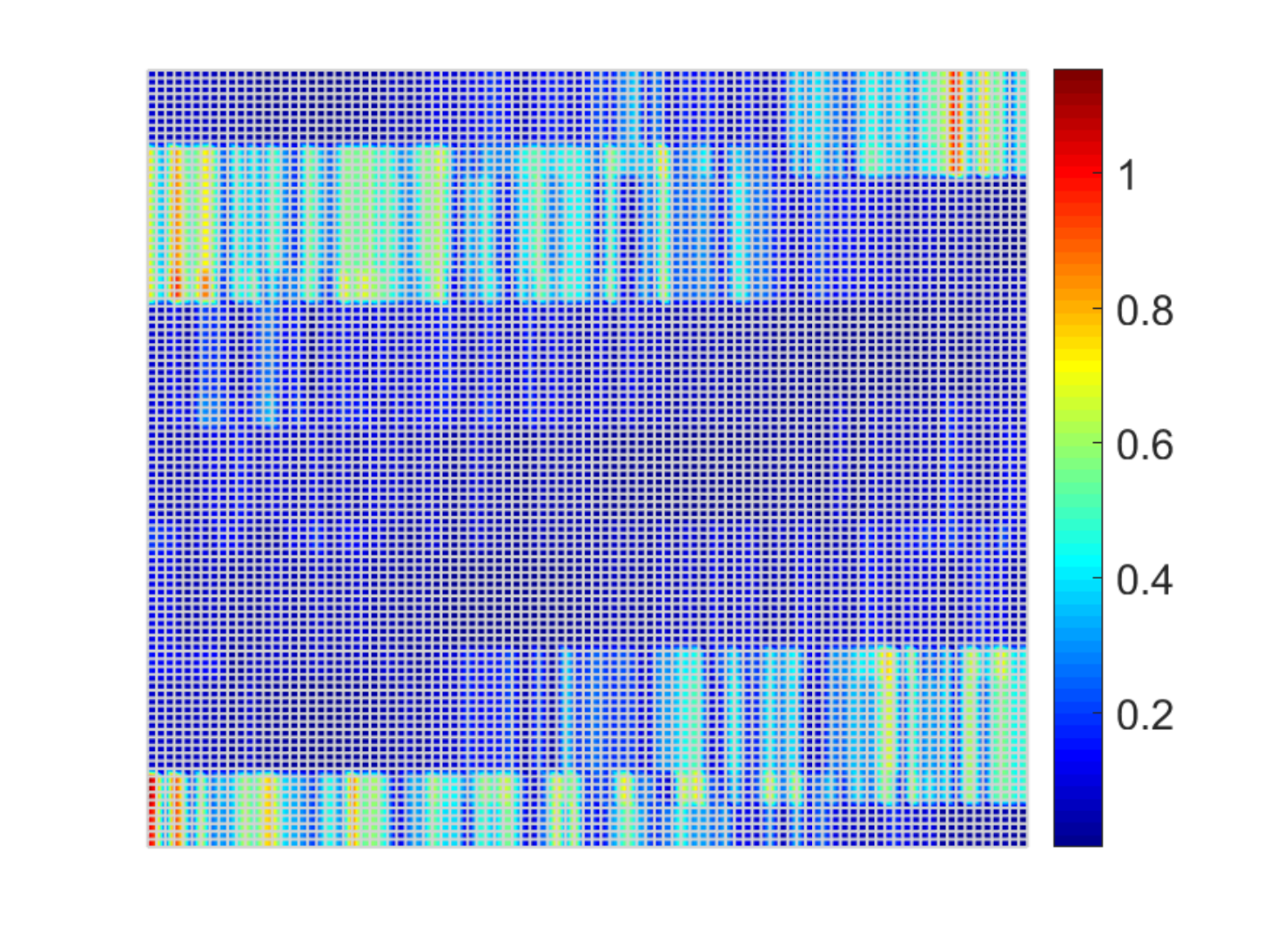}
    \end{minipage}
    }
    \centering
    \caption{ Thermograms of random matrices. (a) local (the first 100 rows and the first 100 columns) thermogram of  $\boldsymbol{A}$ in \eqref{y-grid} in a random experiment. (b) Thermogram of a matrix with each entry randomly drawn from a standard complex Gaussian distribution. (c) local (the first 100 rows and the first 100 columns) thermogram of $\boldsymbol{AF}$ where $\boldsymbol{F}$ is a unitary DFT matrix.}
\end{figure*}

   Recall that the receiver of the considered grant-free satellite-IoT system carriers out joint DAD and CE. With \eqref{y-grid} and the discussions in Subsection D, the joint DAD and CE problem is to  estimate a sparse vector $\boldsymbol{h}$ given the observation $\boldsymbol{y}$. 
Various compressed sensing algorithms have been proposed to solve the linear-inverse problem in the form of \eqref{y-grid}. However, popular Bayesian algorithms, such as AMP and Turbo-CS,
suffer from significant performance degradation when applied to our problem.
This is because AMP and Turbo-CS generally rely on the randomness of the measurement matrix $\boldsymbol{A}$ to ensure convergence. For example, the recovery performance of AMP is guaranteed only when the elements of $\boldsymbol{A}$ are independently and identically distributed (i.i.d.) Gaussian; as for Turbo-CS, $\boldsymbol{A}$ is required to be right-rotationally invariant. When the randomness requirement of $\boldsymbol{A}$ is not met, the recovery  performance of the corresponding algorithms will be seriously degraded. In this work, the structure of  $\boldsymbol{A}$  cannot be arbitrarily designed. From \eqref{ak} and \eqref{Au},  each row of $\boldsymbol{A}$ is the sum of a series of Kronecker products. This  introduces correlation between the elements in a same row of $\boldsymbol{A}$. Fig. \ref{ther1} shows the local thermogram of $\boldsymbol{A}$ in a random experiment, which reflects the amplitude of the elements in $\boldsymbol{A}$. It is clear that the amplitudes of the elements in a same row are correlated, rather than independent and identically distributed. As a comparison, Fig. \ref{ther2} shows the thermogram of a matrix of the same size with each element randomly drawn from the standard complex Gaussian distribution.
Fig. \ref{ther3} shows the local thermogram of $\boldsymbol{AF}$ where $\boldsymbol{F}$ is a discrete Fourier transform (DFT) matrix (which is unitary). Clearly, $\boldsymbol{AF}$ and $\boldsymbol{A}$ show different patterns. Therefore $\boldsymbol{A}$ is not a right-rotationally invariant matrix, and thus does not meet the randomness requirements of both AMP and Turbo-CS. We will see that AMP and Turbo-CS perform poorly for this task in  simulations.

Considering the special structure of $\boldsymbol{A}$ in \eqref{y-grid}, we follow the variance state propagation (VSP) framework proposed in \cite{vsp} and propose the MVSP algorithm to solve this sparse signal recovery problem. Compared with AMP and Turbo-CS, the MVSP algorithm is more robust to the structure of the measurement matrix, which is derived in next section. 

    \section{Receiver Design for Joint DAD and CE}
In this section, we first introduce the probability model and the problem formulation, and then derive  the MVSP  algorithm based on a factor-graph representation of the probability model.

    \subsection{Probability Model}
 For notational convenience, we 
    % define $R=NMU$, $W=KLJ$, and 
    rewrite $\boldsymbol{h}$ as
    \begin{align}
    %   \boldsymbol{h}= \left[h_{1,1}, h_{1,2},\cdots,h_{1,LJ},h_{2,1}, h_{2,2},\cdots,h_{2,LJ},\cdots,h_{K,1}, h_{K,2},\cdots,h_{K,LJ}\right]^T.\nonumber
        \boldsymbol{h}= \left[h_{1,1}, h_{1,2},\cdots,h_{1,LJ},\cdots,h_{K,1}, h_{K,2},\cdots,h_{K,LJ}\right]^T.\nonumber
    \end{align}
    Similarly to \cite{vsp-iccc}, we assign the sparse channel $\boldsymbol{h}$  with a conditional Gaussian prior as 
    \begin{align}\label{eq.24}
        p\left(\boldsymbol{h}|\boldsymbol{v}\right) = \prod_{k=1}^K \prod_{i=1}^{LJ} p\left(h_{k,i}|v_{k,i}\right),
    \end{align}
    where $\boldsymbol{v} = \left[v_{1,1},v_{1,2},\cdots,v_{K,LJ}\right]$, and $p\left(h_{k,i}|v_{k,i}\right) = \mathcal{C}\mathcal{N}\left(h_{k,i};0,v_{k,i}\right)$ is a circularly symmetric complex Gaussian (CSCG) distribution with zero mean and variance $v_{k,i}$, $k\in\left\{1,\cdots,K\right\}$, $i\in\left\{1,\cdots,LJ\right\}$. Each $v_{k,i}$ is assigned with a conditionally independent distribution given by 
     \begin{align}\label{eq.25}
        p\left(v_{k,i}|s_{k,i}\right) = &
            \textup{Gamma}\left(v_{k,i};\gamma_{k,1},\gamma_{k,2}\right) \delta(s_{k,i}-1) +\delta(v_{k,i})\delta(s_{k,i}+1),
    \end{align}
    where  $s_{k,i} \in \{-1, 1\}$ is a hidden binary state;  $\textup{Gamma}\left(v_{k,i};\gamma_{k,1},\gamma_{k,2}\right)$ is the Gamma distribution
    \begin{equation}\label{gamma}
    \textup{Gamma}\!\left(v_{k,i};\gamma_{k,1},\gamma_{k,2}\right)\! =\! \begin{cases}
   \frac{\gamma_{k,2}^{\gamma_{k,1}} v_{k,i}^{\gamma_{k,1}-1} e^{-\gamma_{k,2} v_{k,i}}}{\Gamma\left(\gamma_{k,1}\right)},&\!v_{k,i} > 0,\\
    0, &\!\textup{otherwise},
    \end{cases}
    \end{equation}
    with $\Gamma\left(\gamma_{k,1}\right)=\int_0^{\infty} t^{\gamma_{k,1}-1} e^{-t} {\rm d}t$ being the Gamma function.

\begin{figure}[htbp]
        \centering
        \includegraphics[width=3.2in]{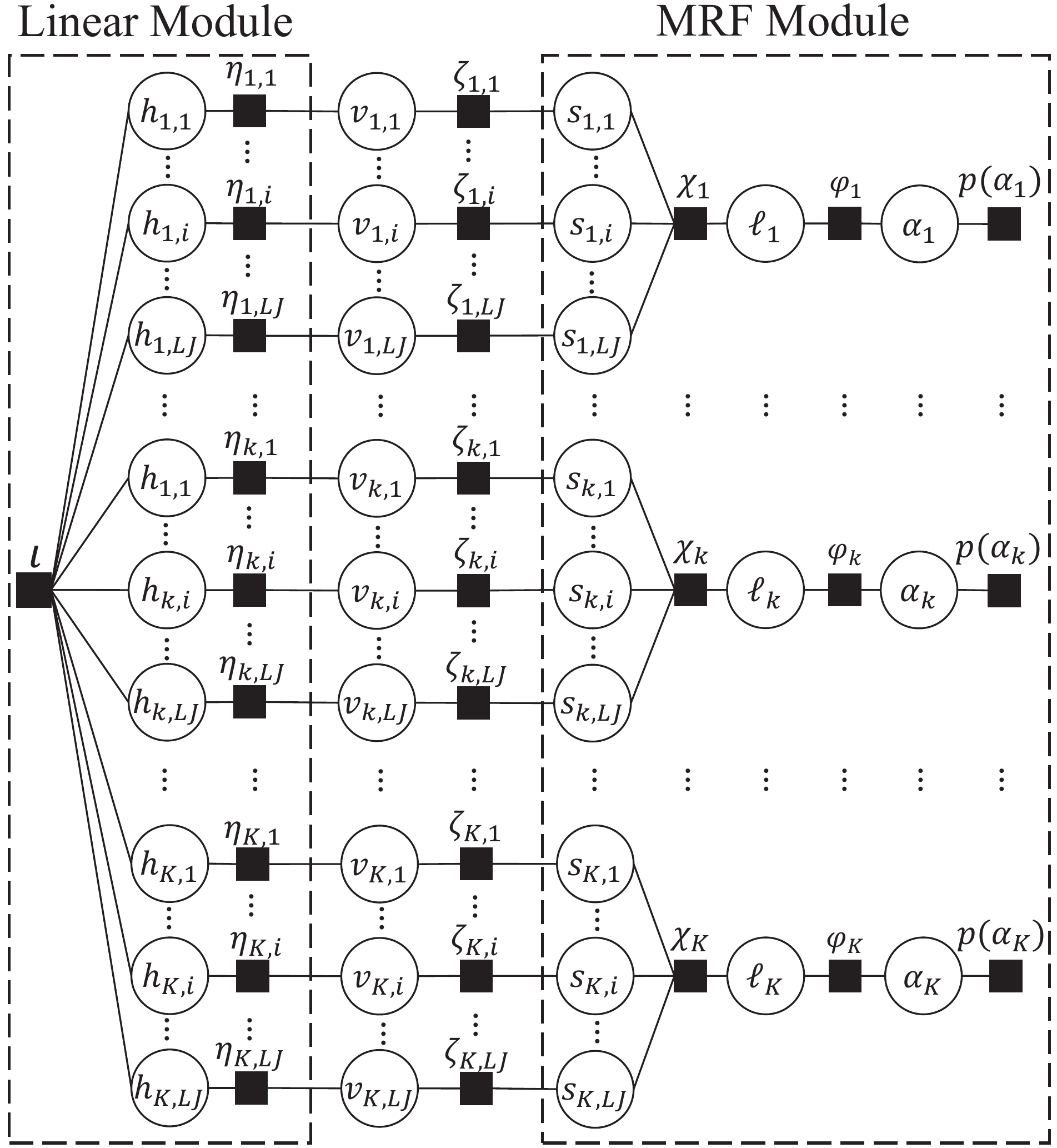}
        \caption{The factor graph characterizing the hierarchical probability model for sparse signals. }
        \label{fg}
    \end{figure}

    Then, a Markov random field (MRF) prior is used to characterize the sparse structure of $\boldsymbol{v}$. The joint probability of the hidden state and device activity variables is modeled as
    \begin{equation}\label{mrf}
        p\left(\boldsymbol{s}_k, \alpha_k\right) \!\propto\! \prod\limits_{i=1}^{LJ}\prod\limits_{z\in \mathcal{D}_i}\left(\varphi\left(s_{k,z},s_{k,i}\right)\right)^{\frac{1}{2}}\!\psi\left(\ell_{k}, \alpha_k\right)p(\alpha_k),
    \end{equation}
    where $\boldsymbol{s}_k = \left[s_{k,1},s_{k,2},\cdots,s_{k,LJ}\right]$, $\varphi\left(s_{k,z},s_{k,i}\right) = \exp\left(\beta s_{k,z}s_{k,i}\right)$, and $\ell_k =\sum_{i}^{LJ}s_{k,i}$; $\mathcal{D}_i$ denotes the set  includes the indexes of the left, right, top and bottom neighbors of $s_{k,i}$, i.e., $\{i-1,i+1,i-J,i+J\}$; 
    % and  $\epsilon$ are and the average number of non-zeros in $\boldsymbol{h}$, respectively
    $\beta$ is the parameter of the MRF, corresponding to the average size of non-zero blocks. Different from the original Ising model \cite{ising}, we add  the constraint $ \psi\left(\ell_{k},\alpha_k\right)$ that represents the conditional probability density of $\ell_k$ given $\alpha_k$. Note that $\ell_k$ is discrete since each $s_{k,i}$ is binary. However, since $LJ$ is large, $\ell_k$ for an active device $k$ (i.e., $\alpha_k=1$) can be well approximated by a Gaussian random variable. Thus, we have
\begin{subequations}
   \begin{equation}\label{psi}
        \psi\left(\ell_{k},\alpha_k\right)\! =\! 
   {\cal{N}}(\ell_k; m_{\psi}, \sigma^2_{\psi})\delta(\alpha_k\! -\! 1)\!+\!
    \delta(l_k\!+\!1)\delta(\alpha_k),
    \end{equation}
    where  $m_{\psi}$ and $\sigma^2_{\psi}$ are the mean and variance of $\ell_k$ conditioned on $\alpha_k\!=\!1$, respectively given by $m_{\psi}=LJ(2\rho_s-1)$ and
    $\sigma^2_{\psi}=4LJ\rho_s(1-\rho_s)$,
    % \begin{align}
    %     % \ell_k &=\sum_{i}s_{k,i}, \\ 
    %     m_{\psi}&=LJ(2\rho_s-1), \\
    %     \sigma^2_{\psi}&=4LJ\rho_s(1-\rho_s),
    % \end{align}
\end{subequations}
         and $\rho_s$ is the sparsity rate of $\{v_{k,i}\}_{i=1}^{LJ}$, i.e., $p(s_{k,i}=1)=\rho_s$.
    The joint probability of $p\left(\boldsymbol{y},\boldsymbol{h},\boldsymbol{v},\boldsymbol{s},\boldsymbol{\alpha}\right)$ can be decomposed as
    \begin{align}\label{eq.30}
        p\left(\boldsymbol{y},\boldsymbol{h},\boldsymbol{v},\boldsymbol{s},\boldsymbol{\alpha}\right) =& p\left(\boldsymbol{y}|\boldsymbol{h}\right)p\left(\boldsymbol{h}|\boldsymbol{v}\right)p\left(\boldsymbol{v}|\boldsymbol{s}\right)p\left(\boldsymbol{s}, \boldsymbol{\alpha}\right)\nonumber\\
        =&p\left(\boldsymbol{y}|\boldsymbol{h}\right)\prod\limits_{k=1}^K \prod\limits_{i=1}^{LJ} p\left(h_{k,i}|v_{k,i}\right)p\left(v_{k,i}|s_{k,i}\right)\prod\limits_{k=1}^K p\left(\boldsymbol{s}_k, \alpha_k\right),
    \end{align}
    where $\boldsymbol{s}=\left[\boldsymbol{s}_1, \cdots,\boldsymbol{s}_K\right]$ and $\boldsymbol{\alpha}=\left[\alpha_1, \cdots,\alpha_K\right]$. 
    The dependencies of the random variables in the factorization \eqref{eq.30} can be shown by a factor graph as depicted in Fig.~\ref{fg}, where circles represent variable nodes and squares represent factor nodes. 
    The factor nodes 
    % $\left\{\zeta_{k,i}\right\}$, $\left\{\eta_{k,i}\right\}$, $\iota$, $\left\{\chi_k\right\}$ and $\{\psi_k\}$
    in Fig.~\ref{fg} are defined as
    \begin{subequations}
    \begin{align}
        \zeta_{k,i}&:p\left(v_{k,i}|s_{k,i}\right),\\
        \eta_{k,i}&:p\left(h_{k,i}|v_{k,i}\right) = \mathcal{C}\mathcal{N}\left(h_{k,i};0,v_{k,i}\right),\\
        \iota &:p\left(\boldsymbol{y}|\boldsymbol{h}\right) = \mathcal{C}\mathcal{N}\left(\boldsymbol{y}-\boldsymbol{A}\boldsymbol{x};\boldsymbol{0},\sigma^2\boldsymbol{I}\right),\\
        \chi_k&:\delta\left(\ell_k-\sum_is_{k,i}\right),\\
        \psi_k&:\psi(\ell_k, \alpha_k).
    \end{align}
    \end{subequations}
    The factor graph in Fig.~\ref{fg} includes two modules, namely, the linear module that handles the linear constraint in \eqref{y-grid} and the MRF module that handles the  MRF prior in \eqref{mrf}.
    
    \subsection{MVSP Algorithm}
    The MVSP algorithm is a sum-product message passing algorithm defined on Fig.~\ref{fg}. A major difference between  MVSP and the original VSP in \cite{vsp} is that variable nodes $\{\ell_k\}$ and $\{\alpha_k\}$ are added to the factor graph for device activity detection, which involves more sophisticated message updates in the MRF module. The details of the MVSP are presented in the following.

    Let $\varpi_{a\to b}$  denote the message passing from node $a$ to node $b$. In Fig.~\ref{fg},   node $v_{k,i}$ receives a message $\varpi_{\eta_{k,i} \to v_{k,i}}$ from  node $\eta_{k,i}$. From the sum-product rule, the message from $v_{k,i}$ to $\zeta_{k,i}$ is still given by $\varpi_{\eta_{k,i} \to v_{k,i}}$. Then the message from $\zeta_{k,i}$ to $s_{k,i}$ is a Bernoulli distribution given by
    \begin{align}\label{eq.36}
        \varpi_{\zeta_{k,i} \to s_{k,i}} \! \propto\! \int_{v_{k,i}} p\left(v_{k,i}|s_{k,i}\right)\varpi_{\eta_{k,i} \to v_{k,i}}
        = \pi_{\zeta_{k,i} \to s_{k,i}}\delta\left(s_{k,i}-1\right)  \!+\!\left(1-\pi_{\zeta_{k,i} \to s_{k,i}}\right)\delta\left(s_{k,i}+1\right),
    \end{align}
    where $\pi_{\zeta_{k,i} \to s_{k,i}}$ is the probability of $s_{k,i} = 1$ specified in the message $\varpi_{\zeta_{k,i} \to s_{k,i}}$.
    \begin{figure}[htbp]
        \centering
        \includegraphics[width=3.4in]{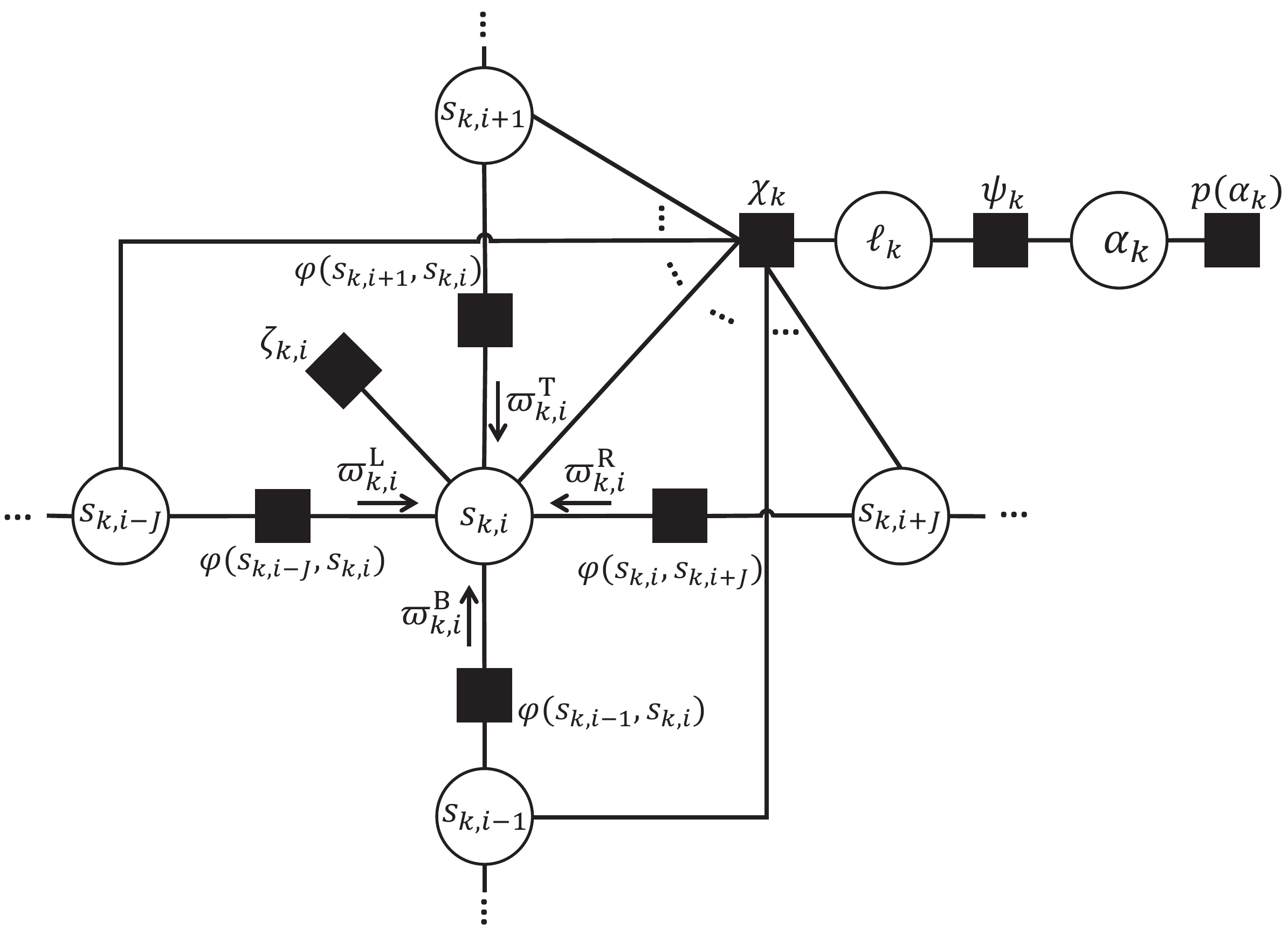}
        \caption{Illustration of the MRF module in MVSP. }
        \label{fig2}
    \end{figure}
    
    We  now  describe the message passing involved in the MRF \eqref{mrf}. A 4-connected MRF is used in MVSP to leverage the block sparsity of each $\boldsymbol{s}_k$ in the delay-Doppler domain. The detailed factor graph characterizing the 4-connected MRF module is  given in Fig.~\ref{fig2}. For clarity, the left, right, top, bottom neighbors to $s_{k,i}$ are reindexed by $\mathcal{I}_i=\left\{i_\text{L}, i_\text{R}, i_\text{T}, i_\text{B}\right\}$ (i.e., $s_{k,i_\text{L}}=s_{k,i-1}$, $s_{k,i_\text{R}}=s_{k,i+1}$, $s_{k,i_\text{T}}=s_{k,i+J}$, $s_{k,i_\text{B}}=s_{k,i-J}$). The left, right, top, and bottom incoming messages of $s_{k,i}$, denoted as $\varpi_{k,i}^\text{L}$, $\varpi_{k,i}^\text{R}$, $\varpi_{k,i}^\text{T}$, and $\varpi_{k,i}^\text{B}$, are Bernoulli distributions. By defining $\mathcal{J}=\left\{\text{L}, \text{R}, \text{T}, \text{B}\right\}$, the incoming message of $s_{k,i}$ from the left is given by
    \begin{subequations}
     \begin{align}\label{mesgl}
        \varpi_{k,i}^\text{L} &\!\propto\! \int_{s_{k,i_\text{L}}} \varpi_{\zeta_{k,i_\text{L}} \to s_{k,i_\text{L}}} \prod\limits_{\jmath\in\mathcal{J}\setminus \text{R}} \varpi^{\jmath}_{k,i_\text{L}} \varphi\left(s_{k,i},s_{k,i_\text{L}} \right)\varpi_{\chi_{k} \to s_{k,i_\text{L}}}\nonumber\\
        &=\lambda_{k,i}^\text{L}\delta\left(s_{k,i}-1\right)+\left(1-\lambda_{k,i}^\text{L}\right)\delta\left(s_{k,i}+1\right),
    \end{align}
    where $\mathcal{J}\setminus \text{R}$ denotes the set $\mathcal{J}$ by excluding the element ``R'',  $\varpi_{\chi_{k} \to s_{k,i_\text{L}}}$ is the message from $\chi_{k}$ to $s_{k,i_\text{L}}$, denoted as
    \begin{align} \label{chi2s}
        \varpi_{\chi_{k} \to s_{k,i_\text{L}}} = & \;\pi_{\chi_{k} \to s_{k,i_\text{L}}} \delta\left(s_{k,i_\text{L}} -1 \right)  + \left(1-\pi_{\chi_{k} \to s_{k,i_\text{L}}}\right)\delta\left(s_{k,i_\text{L}}\ + 1\right),
    \end{align}
    and $\lambda_{k,i}^\text{L}$ is given by \eqref{lambdal}.
    \end{subequations}

    \begin{figure*}[bp] 
        \hrule
      \begin{align}\label{lambdal}
        \lambda_{k,i}^\text{L}\!=\!
        \frac{e^{\beta}\pi_{\zeta_{k,i_\text{L}} \to s_{k,i_\text{L}}}\pi_{\chi_{k} \to s_{k,i_\text{L}}}\!\!\prod\limits_{\jmath\in\mathcal{J}\setminus \text{R}}\!\!\lambda_{k,i_\text{L}}^{\jmath}+e^{-\beta}\left(1\!-\!\pi_{\zeta_{k,i_\text{L}} \to s_{k,i_\text{L}}}\right)\!\!\left(1\!-\!\pi_{\chi_{k} \to s_{k,i_\text{L}}}\right)\prod\limits_{\jmath\in\mathcal{J}\setminus \text{R}}\left(1-\lambda_{k,i_\text{L}}^{\jmath}\right)}
        {\left(e^{\beta}\!+\!e^{-\beta}\right)\!\!\left(\!\pi_{\zeta_{k,i_\text{L}} \to s_{k,i_\text{L}}}\pi_{\chi_{k} \to s_{k,i_\text{L}}} \!\!\prod\limits_{\jmath\in\mathcal{J}\setminus \text{R}} \!\! \lambda_{k,i_\text{L}}^{\jmath} \!+\!\left(1\!-\!\pi_{f_{k,i_\text{L}} \to s_{k,i_\text{L}}}\right)\!\!\left(\!1-\!\pi_{\chi_{k} \to s_{k,i_\text{L}}}\!\right)\!\prod\limits_{\jmath\in\mathcal{J}\setminus \text{R}}\!\left(1-\lambda_{k,i_\text{L}}^{\jmath}\right)\!\!\right)}
    \end{align}
    \end{figure*}
 
    The incoming messages of $s_{k,i}$ from the right, top, and bottom, i.e., $\varpi_{k,i}^\text{R}$, $\varpi_{k,i}^\text{T}$, and $\varpi_{k,i}^\text{B}$, have a form similar to $\varpi_{k,i}^\text{L}$. The message from $\chi_{k}$ to $\ell_k$ is expressed as
    \begin{align}\label{chi2ell}
        \varpi_{\chi_{k} \to \ell_k} \!\propto\!  \int_{\boldsymbol{s}_{k}}\delta\left(\ell_{k}-\sum_{i}s_{k,i}\right)
        \prod_i\left(\varpi_{\zeta_{k,i} \to s_{k,i}}  \prod\limits_{\jmath\in\mathcal{J}} \varpi_{k,i}^{\jmath}  \right)\!.
        % &= \pi_{\psi_{k,i} \to \alpha_k}\delta\left(\alpha_k-1\right)+\left(1-\pi_{\psi_{k,i} \to \alpha_k}\right)\delta\left(\alpha_k\right),
    \end{align}
    Similarly to \eqref{psi}, we approximate \eqref{chi2ell} with a Gaussian distribution:
    \begin{subequations} \label{chi2ell_appro}
       \begin{align}\label{chi2ell2}
        \varpi_{\chi_{k} \to \ell_k} & = {\cal{N}}(\ell_k; m_{\chi_k\to \ell_k}, \sigma_{\chi_k\to \ell_k}^2),
    \end{align}
    where $m_{\chi_k \to \ell_k}$ and $\sigma^2_{\chi_k \to \ell_k}$ are the mean and variance, respectively given by $m_{\chi_k \to \ell_k}  =\sum_i(2\pi_{s_{k,i} \to \chi_k}-1)$ and $\sigma_{\chi_k\to \ell_k}^2  = 4\sum_i\pi_{s_{k,i}}(1-\pi_{s_{k,i}})$, and
     \begin{equation}\label{chi2ell3}
        % m_{\chi_k \to \ell_k} & =\sum_i(2\pi_{s_{k,i} \to \chi_k}-1), \\
        % \sigma_{\chi_k\to \ell_k}^2 & = 4\sum_i\pi_{s_{k,i}}(1-\pi_{s_{k,i}}),\\
        \pi_{s_{k,i}\to \chi_k} \!=\!\frac{\pi_{\zeta_{k,i} \to s_{k,i}} \prod\limits_{\jmath\in {\cal{J}}}\lambda_{k,i}^{\jmath}}{\pi_{\zeta_{k,i} \to s_{k,i}} \!\prod\limits_{\jmath\in {\cal{J}}}\!\lambda_{k,i}^{\jmath} + (1\!-\!\pi_{\zeta_{k,i} \to s_{k,i}}) \prod\limits_{\jmath\in {\cal{J}}}(\!1\!-\lambda_{k,i}^{\jmath})}. 
    \end{equation}
    \end{subequations}
     The message from $\psi_{k}$ to $\alpha_{k}$ is given by
    \begin{subequations} %\label{\psi2alpha}\nonumber\\&
      \begin{align}
        \varpi_{\psi_{k} \to \alpha_{k}}\! &\!\propto\!  \int_{\ell_k} 
        \psi(\ell_k,\alpha_k)  \varpi_{\chi_{k} \to \ell_k}   = \pi_{\psi_{k} \to \alpha_{k}}\delta(\alpha_{k}-1)+(1-\pi_{\psi_{k} \to \alpha_{k}})\delta(\alpha_{k}),\label{psi2alpha1}
    \end{align}
    with $\varpi_{\ell_{k} \to \psi_{k}}\! = \!\varpi_{\chi_{k} \to \ell_{k}}$, where
    \begin{align}
      \pi_{\psi_{k} \to \alpha_{k}} &= \frac{\pi_{\psi_{k} \to \alpha_{k}, 1}}{\pi_{\psi_{k} \to \alpha_{k}, 1} + \pi_{\psi_{k} \to \alpha_{k}, 0}}, \label{psi2alpha2}\\
      \pi_{\psi_{k} \to \alpha_{k}, 1} & = \frac{ e^{-(m_{\chi_k \to \ell_k} - m_{\psi})^2/(2\sigma_{\chi_k \to \ell_k}^2+ 2\sigma_{\psi}^2) } }{\sqrt{2\pi(\sigma^2_{\chi_k \to \ell_k}+\sigma_{\psi}^2)}},\\
      \pi_{\psi_{k} \to \alpha_{k}, 0} & = \frac{e^{-(-1-m_{\chi_k \to \ell_k})^2/2\sigma_{\chi_k \to \ell_k}^2}}{\sqrt{2\pi\sigma^2_{\chi_k \to \ell_k}}}.
    \end{align}
    \end{subequations}
    The message from $\psi_k$ to $\ell_k$ is constant and  given by 
    \begin{align}\label{psi2ell} 
    \varpi_{\psi_k \to \ell_k} & = \int_{\alpha_k} p(\alpha_k) \psi(\ell_k, \alpha_k)  = \rho {\cal{N}}(\ell_k; m_{\psi}, \sigma_{\psi}^2) + (1-\rho)\delta(\ell_k+1).
    \end{align}

    Then we calculate the message from $\chi_{k}$ to $s_{k,i}$ with a similar form of \eqref{chi2s}.
    To obtain $\pi_{\chi_k \to s_{k,i}}$, we first calculate $\lambda_{\chi_k \to s_{k,i}}$ based on the sum-product rule:
      \begin{align}
        \lambda_{\chi_k \to s_{k,i}} =& \int_{\boldsymbol{s}_k\setminus s_{k,i}, \ell_k} \prod_{i'\neq i} \left(\varpi_{\zeta_{k,i'} \to s_{k,i'}} \prod_{\jmath\in{\cal{J}}} \varpi_{k,i'}^\jmath \right)   \varpi_{\psi_{k} \to \ell_{k}}\delta\left(\ell_k-\sum_is_{k,i}\right)\nonumber\\
        =& \int_{ \ell_k}  \!\varpi_{\psi_{k} \to \ell_{k}} \!\int_{\boldsymbol{s}_k\setminus s_{k,i}} \prod_{i'\neq i} \left(\varpi_{\zeta_{k,i'} \to s_{k,i'}} \prod_{\jmath\in{\cal{J}}} \varpi_{k,i'}^\jmath \right)  \delta\left(\ell_k - s_{k,i}-\sum_{i' \neq i}s_{k,i'}\right),
    \end{align}
    where $\boldsymbol{s}_k\setminus s_{k,i}$ denotes all the  entries of  $\boldsymbol{s}_k$  expect  $s_{k,i}$.
    The inner integral is approximated by
    \begin{subequations} \label{eq.33}
    \begin{align}
        {\cal{N}}(\ell_k-s_{k,i}; m_{\chi_k \to s_{k,i}}, \sigma^2_{\chi_k \to s_{k,i}}),
    \end{align}
    where  $m_{\chi_k \to s_{k,i}}$ and $\sigma^2_{\chi_k \to s_{k,i}}$ are  respectively given by
    \begin{align}
        m_{\chi_k \to s_{k,i}} & = \sum_{i' \neq i}(2\pi_{s_{k,i} \to \chi_k}-1), \\
        \sigma^2_{\chi_k \to s_{k,i}} & = 4\sum_{i' \neq i}\pi_{s_{k,i} \to \chi_k}(1- \pi_{s_{k,i} \to \chi_k}).
    \end{align}
    \end{subequations}
    As such, $\lambda_{\chi_k \to s_{k,i}}$ can be given by
        \begin{align} \label{eq.34}
        \lambda_{\chi_k \to s_{k,i}}\! =\! (1-\rho) {\cal{N}}(s_{k,i}; LJ+m_{\chi_k\to s_{k,i}}, \sigma^2_{\chi_k \to s_{k,i}}) \!+ \! \rho {\cal{N}}(s_{k,i}; m_{\chi_k\to s_{k,i}}m_{\psi}, \sigma^2_{\chi_k \to s_{k,i}} \!+\! \sigma_{\psi}^2).
    \end{align}
    Then, $\pi_{\chi_k \to s_{k,i}}$ is given by
    \begin{align} \label{pi-chi2s}
        \pi_{\chi_k \to s_{k,i}}\! =\! \frac{\lambda_{\chi_k \to s_{k,i}}(s_{k,i}=1)}{\lambda_{\chi_k \to s_{k,i}}(s_{k,i}\!=\!1)+\lambda_{\chi_k \to s_{k,i}}(s_{k,i}=\!-1\!)}.
    \end{align}

    The output message of $s_{k,i}$ can be calculated as
    \begin{subequations}
    \begin{align}\label{s2zeta}
        \varpi_{s_{k,i} \to \zeta_{k,i}} = & \pi_{s_{k,i} \to \zeta_{k,i}}\delta\left(s_{k,i}-1\right)+\left(1-\pi_{s_{k,i} \to \zeta_{k,i}}\right)\delta\left(s_{k,i}+1\right),
    \end{align}
    where 
    \begin{align}
        \pi_{s_{k,i} \to \zeta_{k,i}} \!= \!\frac{\pi_{\chi_{k} \to s_{k,i}} \prod\limits_{\jmath\in\mathcal{J}}\lambda_{k,i}^{\jmath}}{\pi_{\chi_k \to s_{k,i}} \!\prod\limits_{\jmath\in\mathcal{J}}\!\lambda_{k,i}^{\jmath} \!+\! (1-\pi_{\chi_{k} \to s_{k,i}}) \!\prod\limits_{\jmath\in\mathcal{J}}\!(1-\lambda_{k,i}^{\jmath})}. 
    \end{align}
    \end{subequations}
    With $\varpi_{s_{k,i} \to \zeta_{k,i}}$, the message from $\zeta_{k,i}$ to $v_{k,i}$ is a Bernoulli-Gamma distribution given by
    \begin{align}\label{zetatov}
        \varpi_{\zeta_{k,i} \to v_{k,i}} \!\! \propto \!\!\!\! \int_{s_{k,i}}\!\!\!\!\! p\left(v_{k,i}|s_{k,i}\right)\varpi_{s_{k,i} \to \zeta_{k,i}}
        \!\!\!= \! \pi_{s_{k,i} \to \zeta_{k,i}}\textup{Gamma}\left(v_{k,i};\gamma_{k,1},\gamma_{k,2}\right)  \!\!+\!\!  \left(1 \!-\!\pi_{s_{k,i} \to \zeta_{k,i}}\right)\!\delta\!\left(v_{k,i}\right).
    \end{align}
    With $\varpi_{v_{k,i} \to \eta_{k,i}} = \varpi_{\zeta_{k,i} \to v_{k,i}}$, the message from $\eta_{k,i}$ to $h_{k,i}$ is given by
    \begin{equation}\label{eq.43}
        \varpi_{\eta_{k,i} \to h_{k,i}} \propto \int_{v_{k,i}} p\left(h_{k,i}|v_{k,i}\right)\varpi_{v_{k,i} \to \eta_{k,i}}.
    \end{equation}
    The message from $h_{k,i}$ to $\iota$ is $\varpi_{h_{k,i} \to \iota} =\varpi_{\eta_{k,i} \to h_{k,i}}$. The message from $\iota$ to  $h_{k,i}$ is
    \begin{align}\label{eq.44}
        \varpi_{\iota \to h_{k,i}} \!\propto\!& \int_{\boldsymbol{h} {\setminus h_{k,i}}} p\left(\boldsymbol{y}|\boldsymbol{h}\right) \prod\limits_{i^{'}\neq i} \varpi_{\eta_{k,i^{'}} \to \iota} \prod\limits_{k^{'}\neq k}\prod\limits_{i''} \varpi_{\eta_{k^{'},i''} \to \iota} \nonumber\\
        =&\int_{\boldsymbol{h} {\setminus h_{k,i}}} \!\!\!\!\! p\left(\boldsymbol{y}|\boldsymbol{h}\right) \prod\limits_{i^{'}\neq i}\int_{v_{k,i^{'}}} \!\!\!\! p(x_{k,i^{'}}|v_{k,i^{'}}) \varpi_{v_{k,i^{'}} \to \eta_{k,i^{'}}} \!\!\!\prod\limits_{k^{'}\neq k}\prod\limits_{i''} \!\! \int_{v_{k^{'},i''}}\!\!\!\!\!\! p(x_{k^{'},i''}|v_{k^{'},i''})\varpi_{v_{k^{'},i''} \to \eta_{k^{'},i''}},
    \end{align}
    where $\boldsymbol{h} {\setminus h_{k,i}}$ denotes all the entries of $\boldsymbol{h}$ except $h_{k,i}$. Clearly, $\varpi_{h_{k,i} \to \eta_{k,i}} = \varpi_{\iota \to h_{k,i}}$. Then, the messages $\varpi_{\eta_{k,i} \to v_{k,i}}$ and $\varpi_{v_{k,i} \to \zeta_{k,i}}$ can be computed as
    \begin{equation}\label{eq.45}
        \varpi_{v_{k,i} \to \zeta_{k,i}} = \varpi_{\eta_{k,i} \to v_{k,i}} \propto \int_{h_{k,i}} p\left(h_{k,i}|v_{k,i}\right) \varpi_{h_{k,i} \to \eta_{k,i}}.
    \end{equation}
    Note that the integrals involved in \eqref{eq.44} and \eqref{eq.45} are difficult to evaluate. From \cite{vsp},
    we can replace the output of the linear module for node $v_{k,i}$ by the mean $\mu_{\eta_{k,i} \to v_{k,i}} = \mathbb{E}_{\varpi_{\eta_{k,i} \to v_{k,i}}}\left[v_{k,i}\right]$. Then $\varpi_{\eta_{k,i} \to v_{k,i}}$ is  approximated by 
    \begin{equation}\label{eq.46}
        \mu_{\eta_{k,i} \to v_{k,i}} = \arg \max\limits_{v_{k,i}} p\left(\boldsymbol{y}|\boldsymbol{v}\right) |_{v_{k,i^{'}}=\mu_{{v_{k,i^{'}} \to \eta_{k,i^{'}}}},\forall i^{'}\neq i},
    \end{equation}
    where $\mu_{v_{k,i} \to \eta_{k,i}} $ is the incoming mean of $v_{k,i}$ for the linear module, i.e., $\mu_{v_{k,i} \to \eta_{k,i}} = \mathbb{E}_{\varpi_{v_{k,i} \to \eta_{k,i}}}\left[v_{k,i}\right]$. Note that $p\left(\boldsymbol{h}|\boldsymbol{y},\boldsymbol{v}\right) \propto p\left(\boldsymbol{y}|\boldsymbol{h}\right)p\left(\boldsymbol{h}|\boldsymbol{v}\right)$  is a complex Gaussian distribution with the mean $\boldsymbol{m}$ and the covariance $\boldsymbol{\Phi}$ given by
    % \begin{align}
    %     \boldsymbol{m} &= \sigma^{-2}\boldsymbol{\Phi}\Bar{\boldsymbol{G}}^H\boldsymbol{y},\label{post-m}\\
    %     \boldsymbol{\Phi} &= \left(\sigma^{-2}\Bar{\boldsymbol{G}}^H\Bar{\boldsymbol{G}}+\boldsymbol{D}^{-1}\right)^{-1},\label{post-psi}
    % \end{align}
    \begin{align}
        \boldsymbol{m} &= \boldsymbol{D}\boldsymbol{A}^H( \sigma^{-2}\boldsymbol{I}+ \boldsymbol{A}\boldsymbol{D} \boldsymbol{A}^H)^{-1}\boldsymbol{y},\label{post-m}\\
        \boldsymbol{\Phi} &= \boldsymbol{D}- \boldsymbol{D}\boldsymbol{A}^H( \sigma^{-2}\boldsymbol{I}+ \boldsymbol{A}\boldsymbol{D} \boldsymbol{A}^H)^{-1}\boldsymbol{A}\boldsymbol{D},\label{post-psi}
    \end{align}
    where $\boldsymbol{D}$ is a diagonal matrix with the $((k-1)LJ+i)$-th diagonal element equals to $\mu_{v_{k,i} \to \eta_{k,i}}$. Hence, by using the ELBO-based method  in \cite{vsp}, the solution of \eqref{eq.46} can be obtained by
    \begin{equation}\label{elbo}
        \mu_{\eta_{k,i} \to v_{k,i}} = |m_{(k-1)LJ + i}|^2 + \phi_{(k-1)LJ + i},
    \end{equation}
    where $m_{(k-1)LJ + i}$ and $\phi_{(k-1)LJ + i}$ are the $\{(k-1)LJ + i\}$-th entry of $\boldsymbol{m}$ and the $\{(k-1)LJ + i\}$-th diagonal element of $\boldsymbol{\Phi}$, respectively. Then $\pi_{\zeta_{k,i} \to s_{k,i}}$ is approximated as 
     \begin{equation}\label{f2s}
        \pi_{\zeta_{k,i} \to s_{k,i}} = \min \left(\frac{\mu_{\eta_{k,i} \to v_{k,i}}}{\gamma_{k,1}/\gamma_{k,2}}, 1\right),
    \end{equation}
   
    Similarly, $\varpi_{v_{k,i} \to \eta_{k,i}}$ can be approximated as
        \begin{equation}\label{v2g}
        \mu_{v_{k,i} \to \eta_{k,i}} = \mathbb{E}_{\varpi_{v_{k,i} \to \eta_{k,i}}}\left[v_{k,i}\right] = \frac{\gamma_{k,1}}{\gamma_{k,2}}\pi_{s_{k,i} \to \zeta_{k,i}}.
    \end{equation} 
    Then, we approximate the mean of the Gamma distribution $\gamma_{k,1}/\gamma_{k,2}$ by $\kappa_k$, where $\kappa_k$ is the mean of the largest $\lfloor3LJ\rho_s\rceil$ elements in $\{\mu_{\eta_{k,i} \to v_{k,i}}\}_{i=1}^{LJ}$. For inactive devices,  this approximation is not accurate because  $\{\mu_{\eta_{k,i} \to v_{k,i}}\}_{i=1}^{LJ}$ has no information about the Gamma distribution. To address this problem, we use the detected active devices' information to help  inactive devices. Let $\kappa_1'\ge\ldots\ge\kappa_K'$ be the reordered sequence of $\kappa_1,\ldots,\kappa_K$, and $K^+=\sum_{k}\hat\alpha_k$  the number of detected active devices, where $\hat\alpha_k$ is given by \eqref{uad}. 
    % Generally, $\kappa_k$ or $\kappa_k'$ indicates the gain of the estimated channel, and the larger it is, the more likely the corresponding devices is  active. 
    We suppose that  the devices corresponding to the largest $\theta_1 K^+$ $\kappa_k'$s  are  active, and their $\kappa_k$s are calculated by $\lfloor3LJ\rho_s\rceil$, where $0<\theta_1<1$. The left $K-\theta_1 K^+$ devices' $\kappa_k$ is approximated by $1/(\theta_2K^+)\sum_{k=1}^{\theta_2K^+}\kappa_k'$, where $\theta_2\ge1$.The detailed derivation of the approximations  in \eqref{eq.46}-\eqref{v2g} can be found in \cite{vsp}. 
    % 
    % We suppose that there are always 
    
    % To obtain a better DAD performance, we cut off $\kappa_k$ and $\kappa_k = \kappa_{\text{thres}}$ if $\kappa_k < \kappa_{\text{thres}}$, where $\kappa_{\text{thres}}$ is the mean of the largest $\lfloor K\rho \rceil$ $\kappa_k$s.
    
    % where $\kappa_{k}$ is to approximate the  mean of Gamma distrubition \eqref{gamma}, i.e., $\gamma_{k,1}/\gamma_{k,2}$, and
    % \begin{align}
    %     \kappa_{k}= \left\{
    %     \begin{array}{lc}
    %          \kappa_{k}', &\kappa_{k}'\ge \kappa_{\text{thres}}\\
    %          \kappa_{\text{thres}}, &\kappa_{k}' < \kappa_{\text{thres}}
    %     \end{array}
    %   \right.
    % \end{align}
    
    % Although the above approximations avoid the the complicated integrals in \eqref{eq.44} and \eqref{eq.45}, $\varpi_{\eta_{k,i} \to v_{k,i}}$ cannot obtained because of these approximations, and thus, $\pi_{\zeta_{k,i} \to s_{k,i}}$ cannot be calculated with \eqref{pi-zetatos}. However, we note that according to \eqref{zetatov}, $\mu_{v_{k,i} \to \eta_{k,i}}$ is given by
    % Inspired by this, we can establish the following map:
    % \begin{equation}
    %     \mu_{\eta_{k,i} \to v_{k,i}} = \frac{\gamma_{k,1}}{\gamma_{k,2}}\pi_{\zeta_{k,i} \to s_{k,i}}.
    % \end{equation}
    % To ensure that $\pi_{\zeta_{k,i} \to s_{k,i}}$ is not greater than 1, we have the following approximation:
    % \begin{equation}
    %     \pi_{\zeta_{k,i} \to s_{k,i}} = \min \left(\frac{\mu_{\eta_{k,i} \to v_{k,i}}}{\gamma_{k,1}/\gamma_{k,2}}, 1\right). \label{f2s}
    % \end{equation}
    % Subsequent message passing can therefore proceed. 
    The above messages are updated iteratively until the algorithm converges. Then, we use the  estimate $\hat{\boldsymbol{h}}=\boldsymbol{m}$ to recover the channel $\hat{\boldsymbol{G}}_u$ based on \eqref{G-hat}, and estimate the device activity as 
    \begin{align} \label{uad}
        \hat\alpha_k=\!\begin{cases}
        1, \quad \text{if}\; \rho \pi_{\psi_k \to \alpha_k}\!>\! (1-\rho)(1-\pi_{\psi_k \to \alpha_k}),\\
        0, \quad \text{if}\; \rho \pi_{\psi_k \to \alpha_k}\!\le\! (1-\rho)(1-\pi_{\psi_k \to \alpha_k}),
        \end{cases} \!\forall k.
    \end{align}
  
    \subsection{Overall Algorithm and Complexity}
    The overall MVSP algorithm is summarized in Algorithm 1. We now analyse the computational complexity of the MVSP algorithm. The complexity in step 5 is $\mathcal{O}(R^3+R^2Q)$.
The complexities in step 8, step 10  and step 11   are all  $\mathcal{O}(Q)$. The complexity in step 13 is  $\mathcal{O}(Q+K)$. Thus, the computational complexity of the MVSP algorithm is dominated by  step 6, and is given by $\mathcal{O}\left(T_{\text{out}}(T_{\text{in1}}(R^3+R^2Q+Q)+K)\right)$.

    % For more details of the VSP algorithm, please refer to [25].
    \begin{algorithm}[htpb] \label{algo1}
        \caption{MVSP algorithm}%算法名字
        \LinesNumbered %要求显示行号
        \KwIn{ $\boldsymbol{y}$, $\boldsymbol{A}$, $T_{\text{out}}$, $T_{\text{in1}}$, $T_{\text{in2}}$}%输入参数
        
        % $\mu_{v_{k,i}\rightarrow \eta_{k,i}}=0, \forall k, \forall i$\; %\;用于换行
        \For{$t_{\text{out}} = 1$ to $T_{\text{out}}$}
        {
             ${\boldsymbol{\hat\mu}}=\left[\mu_{v_{1,1}\rightarrow \eta_{1,1}},\ldots,\mu_{v_{K,LK} \rightarrow \eta_{K,LJ}}\right]^{\text{T}}$ \;
            \For{$t_{\text{in1}}=1$ to $T_{\text{in1}}$}
            {
                $\boldsymbol{D}=\text{diag}(\hat{\boldsymbol{\mu}})$\;
                Compute $\boldsymbol{m}$ and $\boldsymbol{\Phi}$ by using \eqref{post-m} and \eqref{post-psi}\;
                Update $\hat{\boldsymbol{\mu}}=\left[\hat{\mu}_{\eta_{1,1}\rightarrow v_{1,1}},\ldots,\hat{\mu}_{\eta_{K,LJ}\rightarrow v_{K,LJ}}\right]^{\text{T}}$ with \eqref{elbo}\;
            }
           Compute  $\{\pi_{\zeta_{k,i} \rightarrow s_{k,i}}\}$ based on \eqref{f2s} and  $\left[\mu_{\eta_{1,1}\rightarrow v_{1,1}},\ldots,\mu_{\eta_{K,LJ}\rightarrow v_{K, LJ}}\right]^{\text{T}}=\hat{\boldsymbol{\mu}}$\;
            \For{$t_{\text{in2}}=1$ to $T_{\text{in2}}$ }
            {   Compute  $\{\varpi_{k,i}^{\jmath}\}$ based on \eqref{mesgl}\;
                Compute  $\{\pi_{\chi_{k} \rightarrow s_{k,i}}\}$ based on \eqref{eq.33}-\eqref{pi-chi2s}\;
            } 
            Compute  $\{\pi_{\chi_{k} \rightarrow \alpha_{k}}\}$, $\{\hat\alpha_k\}$, and  $\{\mu_{v_{k,i} \rightarrow \eta_{k,i}}\}$ by \eqref{psi2alpha2}, \eqref{uad} and  \eqref{v2g}, respectively\;
        }
        % Compute the \textit{posterior} mean $\boldsymbol{m}$ based on \eqref{post-m} based on $\{\mu_{v_i \rightarrow a_i}\}$\; 
        \KwOut{$\hat{\boldsymbol{h}}=\boldsymbol{m}$, $\hat\alpha_k, \forall k$.}%输出
    \end{algorithm}

    \section{EM-MVSP Algorithm}
    \subsection{Model Mismatch Problem}
    It is worth noting that the true delay and Doppler frequency shift of each path may not fall onto the given grid  $\boldsymbol{\tau}_k$ and $\boldsymbol{\nu}_k$, which results in the model mismatch problem.  To tackle this, we treat $\boldsymbol{\tau}_k$ and $\boldsymbol{\nu}_k$ involved in 
    $\boldsymbol{A}$ as unknown parameters and consider a  parametric dictionary learning method to improve the performance of the MVSP. To this end, we denote $\boldsymbol{A}$ as $\boldsymbol{A}(\boldsymbol{\omega})$ where $\boldsymbol{\omega}=\{\boldsymbol{\tau}_1,\ldots, \boldsymbol{\tau}_{K}, \boldsymbol{\nu}_1,\ldots,\boldsymbol{\nu}_{K}\}$. We learn the parameter set $\boldsymbol{\omega}$ through the EM method. 
    % The EM algorithm is commonly used for parameter estimation in probability models with hidden variables. 

    \subsection{EM Framework}
Each iteration of the EM method consists of two steps: 
\begin{itemize}
    \item E-step: To calculate the posterior distribution of hidden variables based on the estimates obtained in the previous iteration, and then formulate a function $\mathcal{Q}$ which is the posterior mean of the log-likelihood function with respect to the hidden variables; 
    \item M-step: To maximize  $\mathcal{Q}$  with respect to the hidden variables to  update parameters. 
\end{itemize}
E-Step and M-step iterate until converge. In our problem, the observed variables are  the received signal $\boldsymbol{y}$ in \eqref{y-grid}, the hidden variables are the channel representation vector $\boldsymbol{h}$, and the parameters to be estimated are $\boldsymbol{\omega}$. Note that the MVSP algorithm can provide approximate  posterior distributions of $\boldsymbol{h}$ in the iteration. Therefore, in this problem, the key to update $\boldsymbol{\omega}$ is the M-step, i.e., to maximize the posterior mean of the log-likelihood function $\mathcal{Q}$. Denote by $\boldsymbol{\omega}^{(i)}$  the parameter update obtained in the $i$-th iteration. Then the  $\mathcal{Q}$ function is formulated as
\begin{align} 
    \mathcal{Q}\left(\boldsymbol{\omega}, \boldsymbol{\omega}^{\left(i\right)}\right)=&\int_{\boldsymbol{h}} \text{ln} p\left(\boldsymbol{y},\boldsymbol{h}|\boldsymbol{\omega}\right)p\left(\boldsymbol{h}|\boldsymbol{y},\boldsymbol{\omega}^{\left(i\right)}\right) \nonumber\\
    =&\int_{\boldsymbol{h}} \text{ln}p\left(\boldsymbol{y}|\boldsymbol{h},\boldsymbol{\omega}\right)p\left(\boldsymbol{h}|\boldsymbol{y},\boldsymbol{\omega}^{\left(i\right)}\right) \int_{\boldsymbol{h}} \text{ln}p\left(\boldsymbol{h}\right)p\left(\boldsymbol{h}|\boldsymbol{y},\boldsymbol{\omega}^{\left(i\right)}\right).\label{Qfunc}
\end{align}
Note that the second term in \eqref{Qfunc} is irrelevant to $\boldsymbol{\omega}$. Therefore, M-step is equivalent to minimizing
\begin{align} \label{Ffunc}
    \mathcal{F}\left(\boldsymbol{\omega}, \boldsymbol{\omega}^{\left(i\right)}\right)=-\int_{\boldsymbol{h}} \text{ln}p\left(\boldsymbol{y}|\boldsymbol{h},\boldsymbol{\omega}\right)p\left(\boldsymbol{h}|\boldsymbol{y},\boldsymbol{\omega}^{\left(i\right)}\right).
\end{align}
From the additive CSCG noise model in \eqref{y-grid}, we have
\begin{align} \label{posty}
    p\left(\boldsymbol{y}|\boldsymbol{h},\boldsymbol{\omega}\right) = \mathcal{CN}\left(\boldsymbol{y}-\boldsymbol{A}\left(\boldsymbol{\omega}\right)\boldsymbol{h}; \boldsymbol{0}, \sigma^2\boldsymbol{I}\right).
\end{align}
In each outer iteration of the MVSP algorithm, the posterior
distribution of the channel representation $\boldsymbol{h}$ given by the linear
module is approximated by
\begin{align} \label{posth}
    p\left(\boldsymbol{h}|\boldsymbol{y},\boldsymbol{\omega}^{\left(i\right)}\right) = \mathcal{CN}\left(\boldsymbol{h}-\boldsymbol{m}\left(\boldsymbol{\omega}^{\left(i\right)}\right); \boldsymbol{0},\boldsymbol{\Phi}\left(\boldsymbol{\omega}^{\left(i\right)}\right)\right),
\end{align}
where $\boldsymbol{m}\left(\boldsymbol{\omega}^{\left(i\right)}\right)$ and $\boldsymbol{\Phi}\left(\boldsymbol{\omega}^{\left(i\right)}\right)$ are the posterior mean and the covariance matrix of $\boldsymbol{h}$ calculated based on $\boldsymbol{y}$ and $\boldsymbol{A}\left(\boldsymbol{\omega}^{\left(i\right)}\right)$. Plugging
\eqref{posty} and \eqref{posth} into \eqref{Ffunc}, we have
\begin{align} \label{Ffunc2}
    \mathcal{F}\left(\boldsymbol{\omega}, \boldsymbol{\omega}^{\left(i\right)}\right) = & \frac{1}{{\sigma^2}}\int_{\boldsymbol{h}}
    \Big(\boldsymbol{y}^{\text{H}}\boldsymbol{y} -  \boldsymbol{y}^{\text{H}}\boldsymbol{A}(\boldsymbol{\omega})\boldsymbol{h} - \boldsymbol{y}^{\text{H}} \boldsymbol{A}^{\text{H}}(\boldsymbol{\omega}) \boldsymbol{y}  \nonumber \\
    & +   \boldsymbol{h}^{\text{H}}\boldsymbol{A}^{\text{H}}(\boldsymbol{\omega})\boldsymbol{A}\left(\boldsymbol{\omega}\right)\boldsymbol{h} \Big)
    p\left(\boldsymbol{h}|\boldsymbol{y},\boldsymbol{\omega}^{\left(i\right)}\right) + \text{ln}(\pi\sigma^2). 
\end{align}
Remove the irrelevant terms in \eqref{Ffunc2} and define a new objective
function
\begin{align} \label{Gfunc2}
    \mathcal{G}\left(\boldsymbol{\omega}, \boldsymbol{\omega}^{\left(i\right)}\right)=& \int_{\boldsymbol{h}} \Big( -  \boldsymbol{y}^{\text{H}}\boldsymbol{A}\left(\boldsymbol{\omega}\right)\boldsymbol{h} - \boldsymbol{h}^{\text{H}} \boldsymbol{A}^{\text{H}}\left(\boldsymbol{\omega}\right) \boldsymbol{y} + \boldsymbol{h}^{\text{H}}\boldsymbol{A}^{\text{H}}\left(\boldsymbol{\omega}\right)\boldsymbol{A}\left(\boldsymbol{\omega}\right)\boldsymbol{h} \Big) p\left(\boldsymbol{h}|\boldsymbol{y},\boldsymbol{\omega}^{\left(i\right)}\right) \nonumber \\
    =&-2\mathcal{R}\left(\boldsymbol{y}^{\text{H}} \boldsymbol{A}\left(\boldsymbol{\omega}\right) {\boldsymbol{m}}\left(\boldsymbol{\omega}^{\left(i\right)}\right)\right)  + \text{Tr}\left(\boldsymbol{A}^{\text{H}}\left(\boldsymbol{\omega}\right) \boldsymbol{A}\left(\boldsymbol{\omega}\right) \int_{\boldsymbol{h}}\boldsymbol{h}\boldsymbol{h}^{\text{H}}p\left(\boldsymbol{h}|\boldsymbol{y},\boldsymbol{\omega}^{\left(i\right)}\right) \right)\nonumber \\
    =& -2\mathcal{R}\left(\boldsymbol{y}^{\text{H}} \boldsymbol{A}\left(\boldsymbol{\omega}\right) \boldsymbol{m}\left(\boldsymbol{\omega}^{(i)}\right)\right)  + \text{Tr}\left(\boldsymbol{A}^{\text{H}}\left(\boldsymbol{\omega}\right) \boldsymbol{A}\left(\boldsymbol{\omega}\right) \boldsymbol{\Omega}^{\left(i\right)} \right),
\end{align}
% Then, the objective function  $\mathcal{G}\left(\boldsymbol{\omega}, \boldsymbol{\omega}^{\left(i\right)}\right)$ can be reorganized as
% \begin{align} \label{Gfunc2}
%     \mathcal{G}\left(\boldsymbol{\omega}, \boldsymbol{\omega}^{\left(i\right)}\right)=
% \end{align}
where $\boldsymbol{\Omega}^{\left(i\right)}= \boldsymbol{\Phi}\left(\boldsymbol{\omega}^{\left(i\right)}\right)+\boldsymbol{m} \left( \boldsymbol{\omega}^{\left(i\right)} \right) \boldsymbol{m}^{\text{H}}\left(\boldsymbol{\omega}^{\left(i\right)}\right)$.

In summary, in each update of $\boldsymbol{\omega}$ through  EM, our method is to calculate the posterior distribution of $\boldsymbol{h}$ and construct the objective function $\mathcal{G}\left(\boldsymbol{\omega}, \boldsymbol{\omega}^{\left(i\right)}\right)$ based on $\boldsymbol{y}$ and $\boldsymbol{\Omega}^{(i)}$. It is not easy to obtain an analytical solution to this problem with respect to both $\boldsymbol{\tau}_k$ and $\boldsymbol{\nu}_k$. Therefore, we use the gradient descent method to  minimize $\mathcal{G}\left(\boldsymbol{\omega}, \boldsymbol{\omega}^{\left(i\right)}\right)$ and  then update
$\boldsymbol{\omega}$  in each iteration. 

 \begin{algorithm}[htbp]
    \caption{EM-MVSP algorithm}
    \LinesNumbered 
    \KwIn{ $\boldsymbol{y}$, $\{\boldsymbol{x}_u\}$, $T_{\text{out}}$, $T_{\text{in}}$, $i_{\text{max}}$.}
    
    Initialization: $i=0, \boldsymbol{\omega}^{\left(0\right)}=\{\boldsymbol{\tau}_1^{\left(0\right)},\ldots, \boldsymbol{\tau}_{K}^{\left(0\right)}, \boldsymbol{\nu}_1^{\left(0\right)},\ldots,\boldsymbol{\nu}_{K}^{\left(0\right)}\}$\; 
    \While{the stopping criterion is not met}
    {
        Generate  $\boldsymbol{A}\left(\boldsymbol{\omega}^{\left(i\right)}\right)$ based on   $\boldsymbol{\omega}^{\left(i\right)}$ and $\{\boldsymbol{x}_u\}$\;
        With $\boldsymbol{y}$ and $\boldsymbol{A}\left(\boldsymbol{\omega}^{\left(i\right)}\right)$, calculate  $\boldsymbol{m}\left(\boldsymbol{\omega}^{\left(i\right)}\right)$,  $\boldsymbol{\Phi}\left(\boldsymbol{\omega}^{\left(i\right)}\right)$  and  $\hat\alpha_k\left(\boldsymbol{\omega}^{\left(i\right)}\right)$ by Algorithm \ref{algo1}\;
        Minimize \eqref{Gfunc2} with respect to $\boldsymbol{\tau}_k$, and obtain the updated $\boldsymbol{\tau}^{\left(i+1\right)}_k$, $\forall k$\;
        Minimize \eqref{Gfunc2} with respect to $\boldsymbol{\nu}_k$, and obtain the updated $\boldsymbol{\nu}^{\left(i+1\right)}_k$, $\forall k$\;
      $\boldsymbol{\omega}^{\left(i+1\right)}=\{\boldsymbol{\tau}_1^{\left(i+1\right)},\ldots, \boldsymbol{\tau}_{K}^{\left(i+1\right)}, \boldsymbol{\nu}_1^{\left(i+1\right)},\ldots,\boldsymbol{\nu}_{K}^{\left(i+1\right)}\}$, $i=i+1$\;
      
    }  
    \KwOut{$\hat{\boldsymbol{h}}={\boldsymbol{m}}\left(\boldsymbol{\omega}^{(i)}\right)$, $\hat\alpha_k=\hat\alpha_k\left(\boldsymbol{\omega}^{(i)}\right)$ .}
\end{algorithm}
\subsection{Overall Algorithm and Complexity}

Algorithm 2  summarizes    the EM-MVSP algorithm. The stopping criterion of the iteration is generally set to ``$\|\boldsymbol{\omega}^{\left(i+1\right)}-\boldsymbol{\omega}^{\left(i\right)}\|^2$ is less than a small positive threshold or $i\ge i_{\text{max}}$'', where $i_{\text{max}}$ is the maximal number of iterations of the EM. We now analyse the computational complexity of the EM-MVSP algorithm. The complexity in step 3 is $\mathcal{O}\left(RQ\left(M+LJ\right)\right)$. The complexities in step 5 and step 6 are both  $\mathcal{O}\left(Q^3+Q^2R\right)$. As mentioned in Section \uppercase\expandafter{\romannumeral4}-C, the complexity in step 4 is $\mathcal{O}(T_{\text{out}}T_{\text{in1}}(R^3+R^2Q))$. Thus, the computational complexity of the EM-MVSP algorithm is given by
$\mathcal{O}\left(i_{\text{max}}\left(Q^3+Q^2R+T_{\text{out}}(T_{\text{in1}}(R^3+R^2Q+Q)+K)\right)\right)$.

\section{Simulation Results}
\begin{figure}[h]
        \centering
        \includegraphics[width=1.8in]{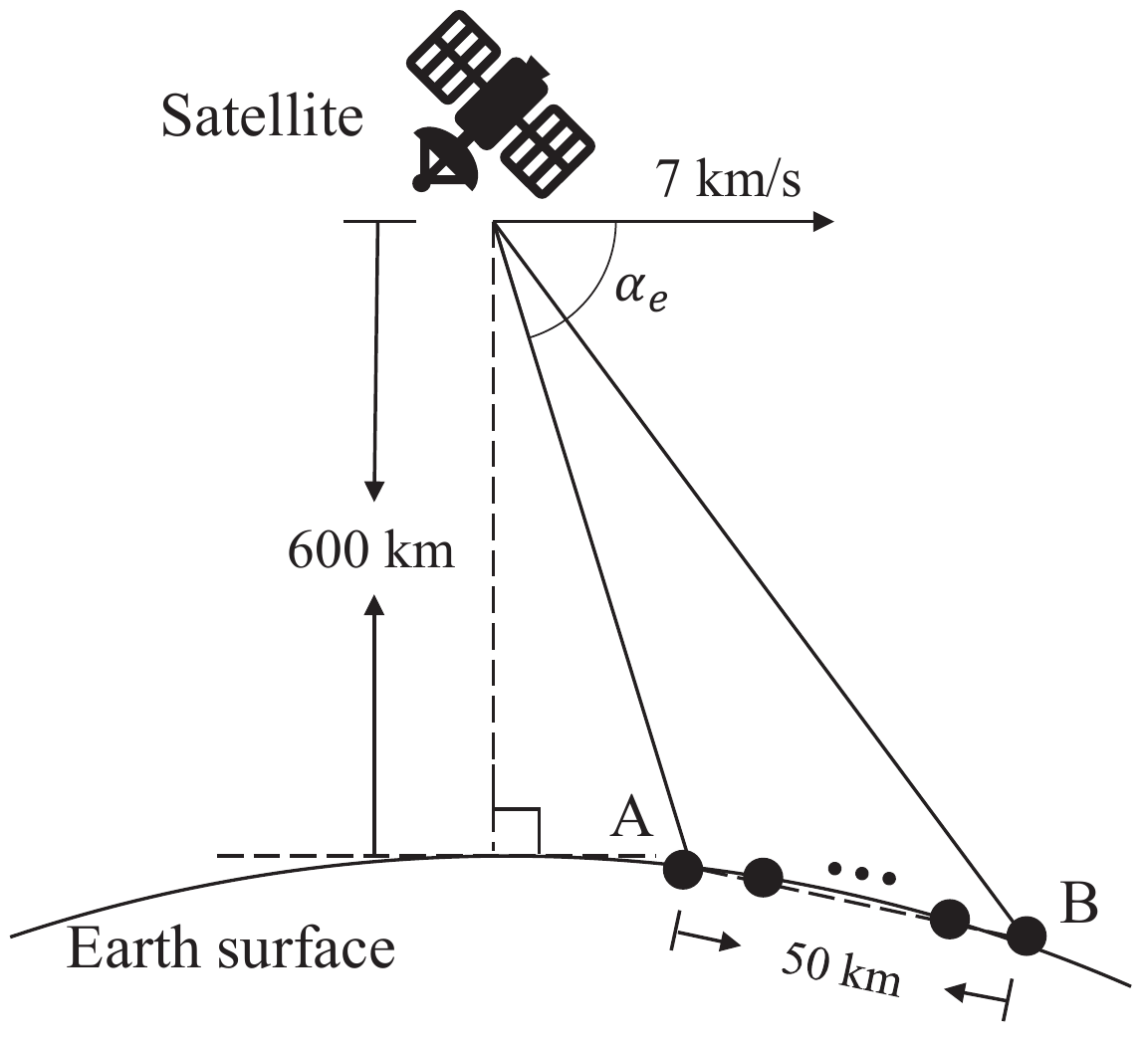}
        \caption{Illustration of the simulated satellite-IoT system.}
        \label{beamsim}
    \end{figure}

In this section, we carry out simulations to demonstrate
the effectiveness of the proposed algorithms.  
The CE  and DAD performance are respectively measured in terms of  NMSE and  detection error
probability  $P_e= \mathbb{E}[1/K\sum_k ( p(\hat{\alpha}_k=0|\alpha_k=1) + p(\hat{\alpha}_k=1|\alpha_k=0) )]$ over 100 independent trials.  The signal-noise-ratio (SNR) is defined as $\sum_u\|\boldsymbol{G}_u\boldsymbol{x}_u\|^2/(R\sigma^2)$. The baseline methods used for comparison includes OMP \cite{OMP}, GAMP \cite{GAMP}, structured TCS (STCS) \cite{STCS}, SBL \cite{SBL},  and pattern-coupled SBL (PCSBL) \cite{PCSBL}. The iteration number of EM in EM-MVSP is 4.   $\{x_{k,m,u}\}$ are generated from  the standard complex Gaussian distribution. %${\cal{CN}}(0, 1)$
The  parameters in the simulation are listed as follows.  
The satellite operates at a speed of 7 km/s in an orbit of 600 km,  and the coverage  of  the beam is a circle with a diameter of 50 km. As shown in Fig.~\ref{beamsim}, device A  and device B  are at the border of the beam coverage, and the elevation angle between the satellite and device  A is $\alpha_e$. 
% Thus, the differential Doppler frequency shift  between device 1 and device 2 are about 12 kHz,  which are assumed to be the maximal  differential Doppler frequency shift $\nu_{\text{max}}$ in the beam after compensation. The maximal differential delay is set to $\tau_{\text{max}}=9 \mu$s after compensation. 
% The size of the delay-Doppler grid is $L=6$ and $J=20$, while the grid values of $\boldsymbol{\tau}_k$  are uniformly selected  from $0$ to $\tau_{\text{max}}$, and the grid values of $\boldsymbol{\nu}_k$  are  uniformly selected  from $-\nu_{\text{max}}/2$ kHz to $\nu_{\text{max}}/2$ kHz. In other words, the grid spacing is 1.5 $\mu$s and 750 Hz, respectively.  
The velocity of  devices follows the uniform distribution with [0, 120] km/h,  the delay spreading is 0.5 $\mu$s, the length of CP  is $T_{\text{\text{cp}}}= 2 \;\mu\text{s}$. We adopt the TDL-A channel model in \cite{3gpp}. 
The large scale fading is  compensated since it changes slightly over the devices in a beam. 
The total number of OFDM symbols in a transmission frame is $UN=12$, and  the subcarrier spacing is $\Delta f = 15$ kHz. 
% There are $K=100$ potential devices in the beam and the sparsity rate is $\rho=0.1$.

\begin{figure}[t]
\centering
\subfigure[CE performance]{
\begin{minipage}[t]{0.5\linewidth}
\centering
\includegraphics[width=2.4in]{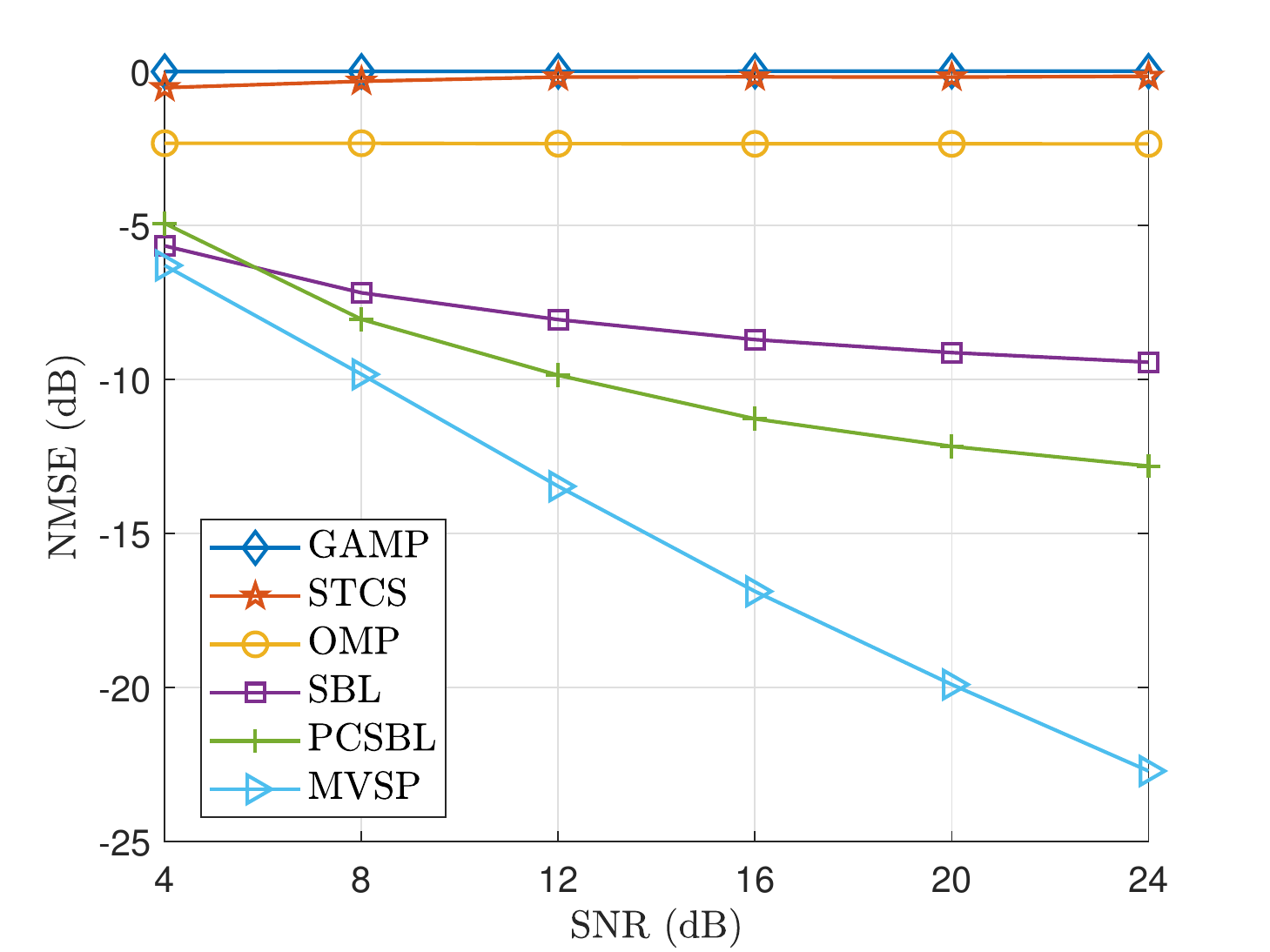}
\label{on-nmse}
\end{minipage}
}\subfigure[DAD performance]{
\begin{minipage}[t]{0.5\linewidth}
\centering
\includegraphics[width=2.4in]{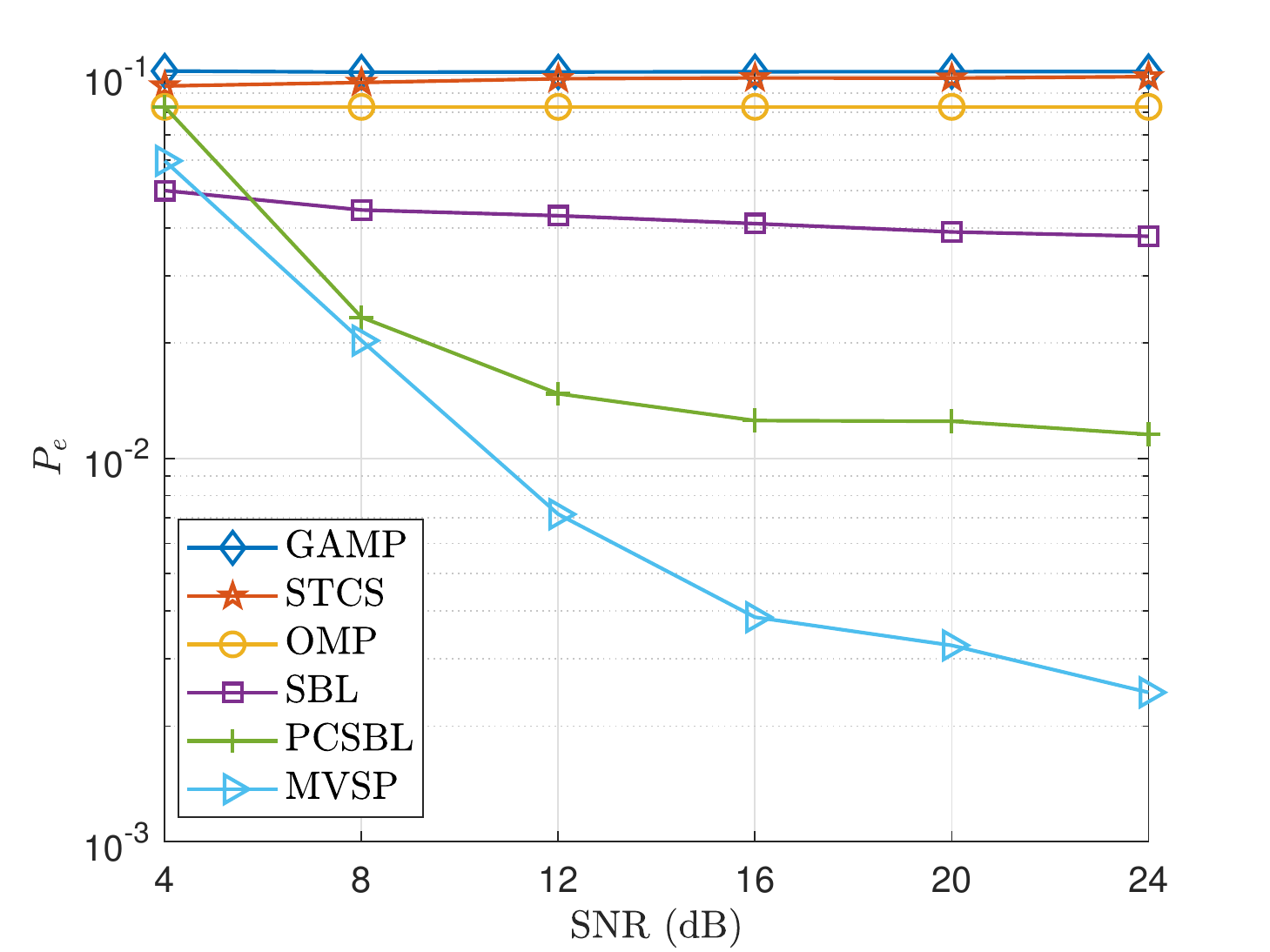}
\label{on-perr}
\end{minipage}
}  
\caption{CE and DAD performances under various SNR and on-grid settings. Related parameters are $K=200$, $\rho = 0.1$, $N=4$,  $L=4$, $J=24$, $M=32$ and $\alpha_e = 50^\circ$.}
\end{figure}

We first consider the scenario when there is no mismatch between the real channel and the grid-based model, that is, the delay and the Doppler frequency shift of each channel path fall onto the  delay-Doppler grid. 
In Fig. \ref{on-nmse}  we show the CE NMSE  versus the SNR. The proposed MVSP outperforms other baseline methods as SNR increases. In particular, due to the special structure of the measurement matrix discussed in Section \ref{PD}, we notice that GAMP and STCS behave poorly in this task, with a NMSE about 0 dB.  PCSBL  outperforms SBL but still has a large performance gap with the proposed MVSP algorithm. In the SNR range of $[16,24]$ dB,  the MVSP algorithm is at least 5 dB better than other baseline methods in NMSE.
Fig. \ref{on-perr} shows the detection error probability versus the SNR.
We see that as the SNR increases, the $P_e$ of GAMP, STCS, OMP and SBL can hardly be improved. MVSP obviously outperform the baseline methods, especially when the SNR is over 16 dB. 
%%% on

%%% n
\begin{figure}[t]
\centering
\subfigure[CE  performance]{
\begin{minipage}[t]{0.5\linewidth}
\centering
\includegraphics[width=2.4in]{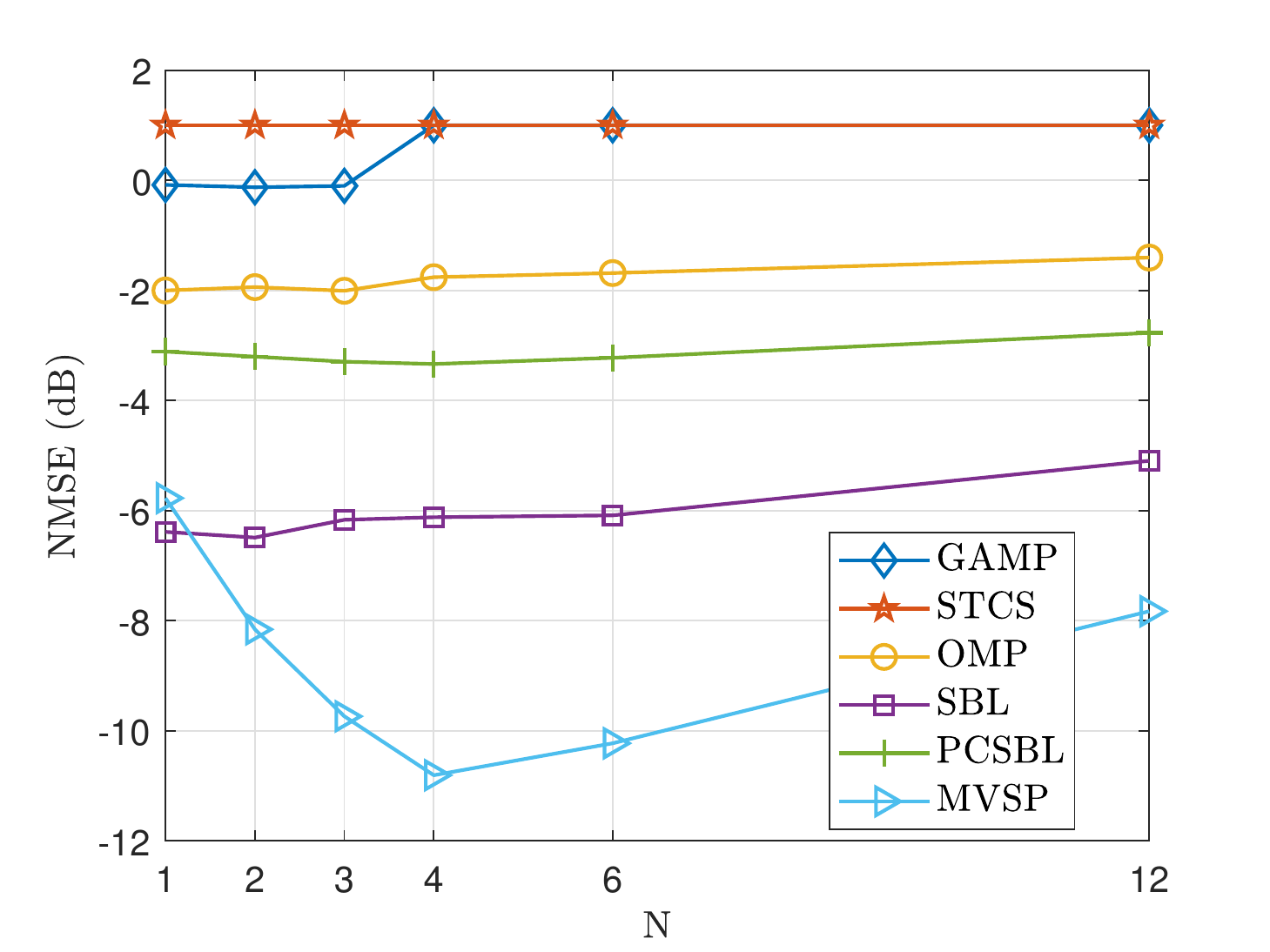}
\label{off-n-nmse}
\end{minipage}
}\subfigure[DAD performance]{
\begin{minipage}[t]{0.5\linewidth}
\centering
\includegraphics[width=2.4in]{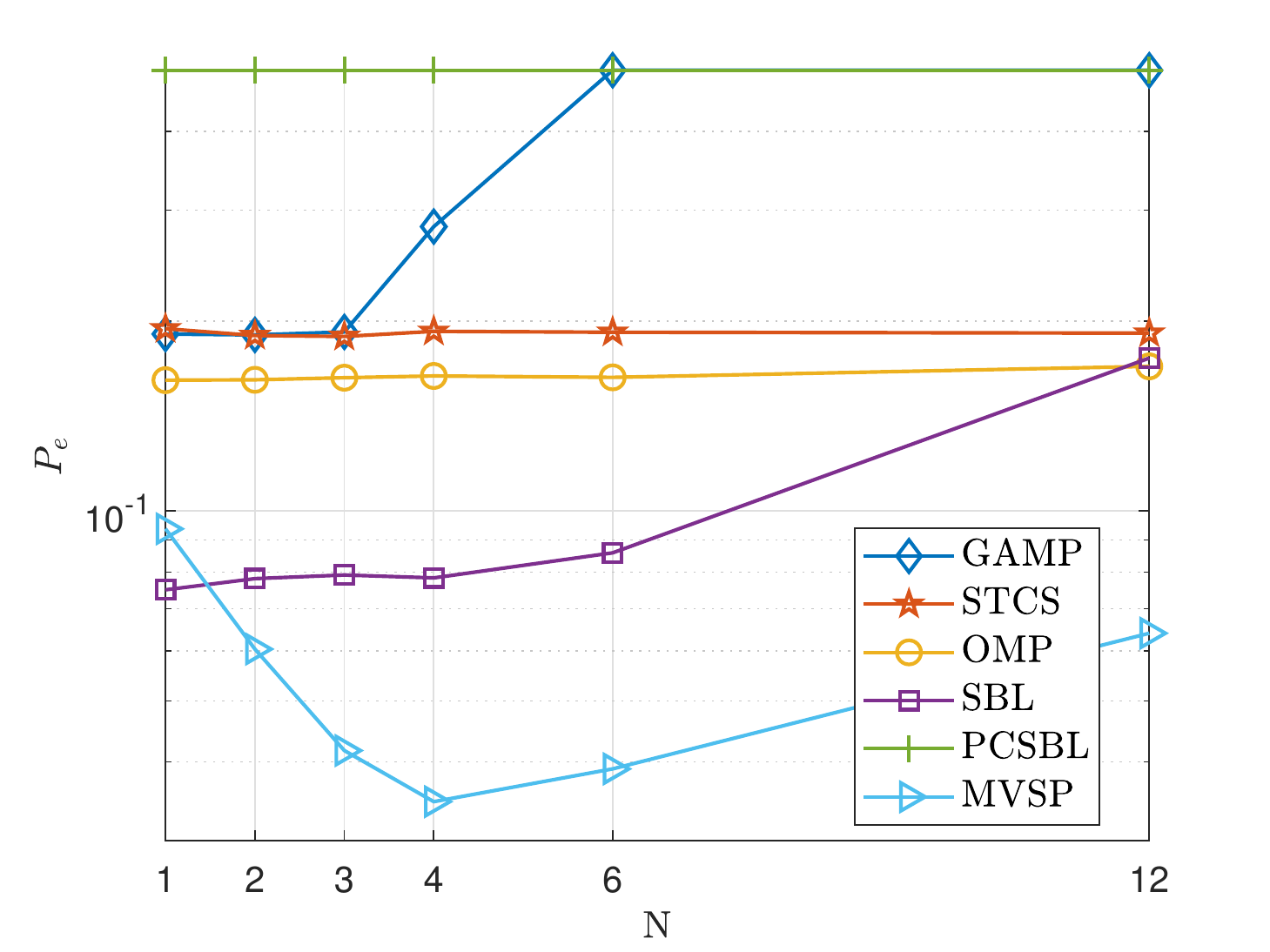}
\label{off-n-perr}
\end{minipage}
}  
\caption{CE and DAD performances under various number of cascaded OFDM symbols  $N$ and off-grid settings. Related parameters are $K=100$, $\rho = 0.2$,  $L=6$, $J=20$, SNR=20 dB, $M=16$ and $\alpha_e = 70^\circ$. }
\end{figure}

%%% 50
\begin{figure}[t]
\centering
\subfigure[CE  performance]{
\begin{minipage}[t]{0.5\linewidth}
\centering
\includegraphics[width=2.4in]{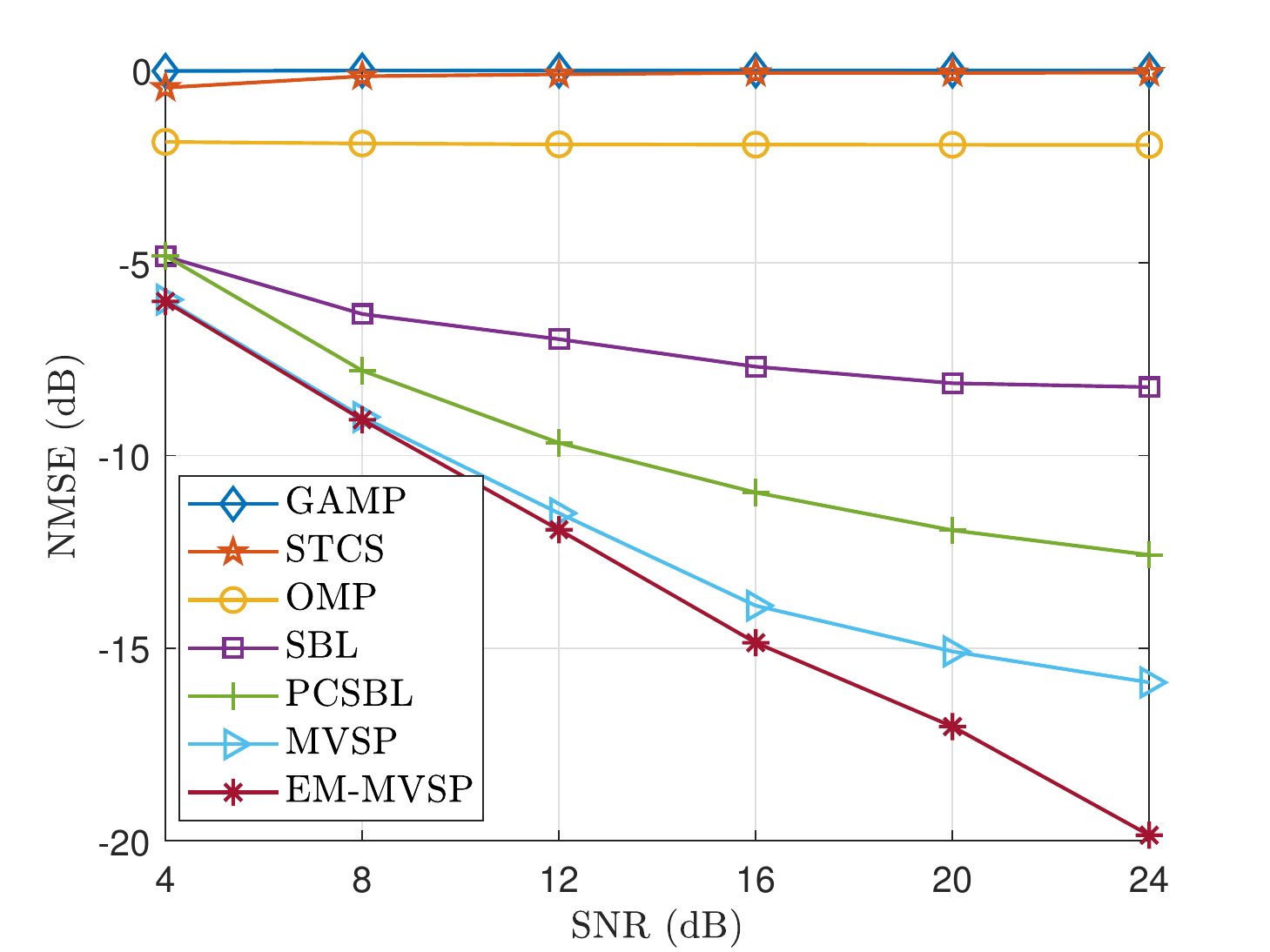}
\label{off-snr-nmse}
\end{minipage}
}\subfigure[DAD performance]{
\begin{minipage}[t]{0.5\linewidth}
\centering
\includegraphics[width=2.4in]{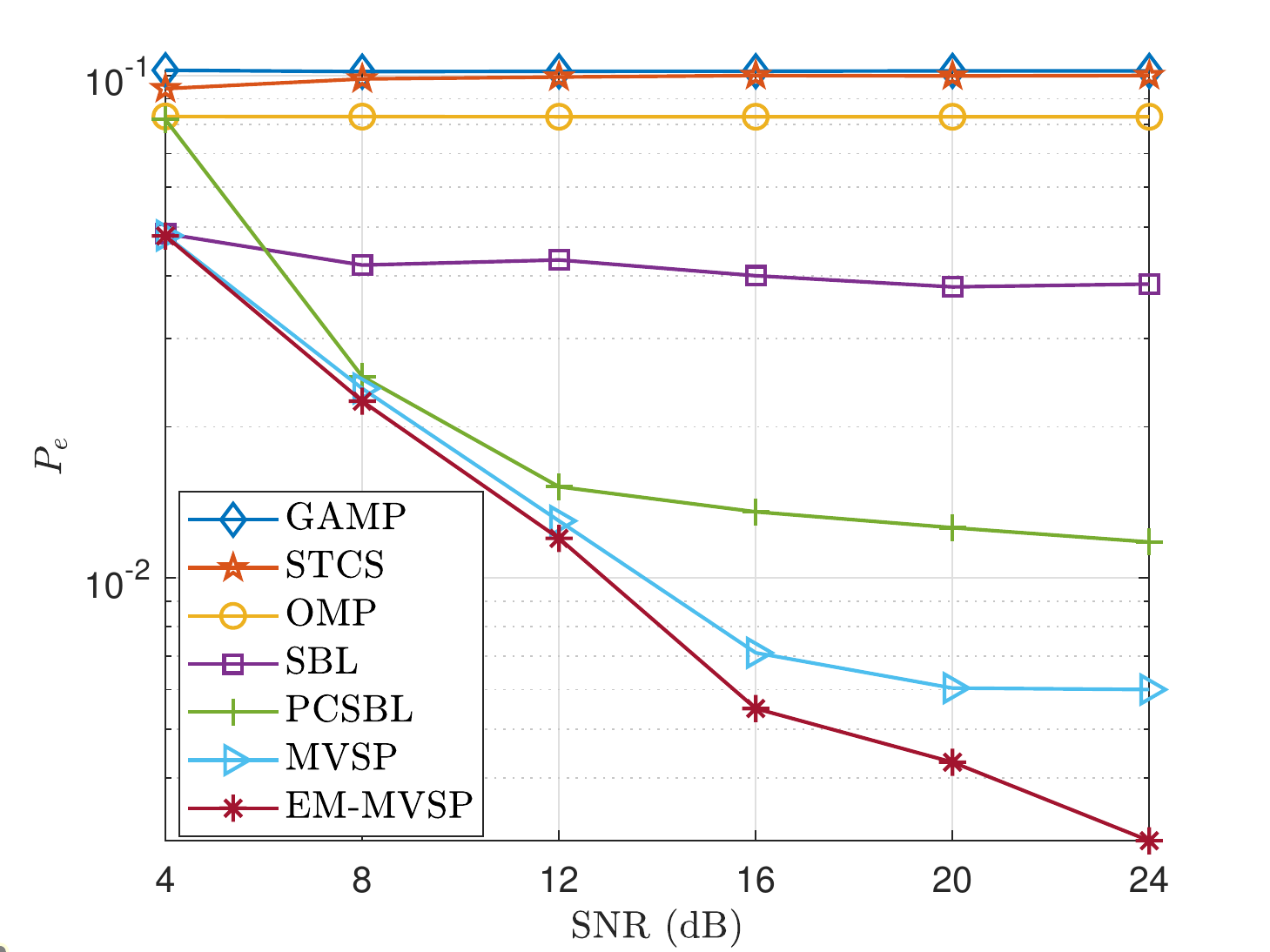}
\label{off-snr-perr}
\end{minipage}
}  
\caption{CE and DAD performances under various SNR and off-grid settings, and $\alpha_e = 50^\circ$. Related parameters are $K=200$, $\rho = 0.1$, $N=4$,  $L=4$, $J=24$ and $M=32$. }
\label{off-50}
\end{figure}

%%% 90
\begin{figure}[t]
\centering
\subfigure[CE  performance]{
\begin{minipage}[t]{0.5\linewidth}
\centering
\includegraphics[width=2.4in]{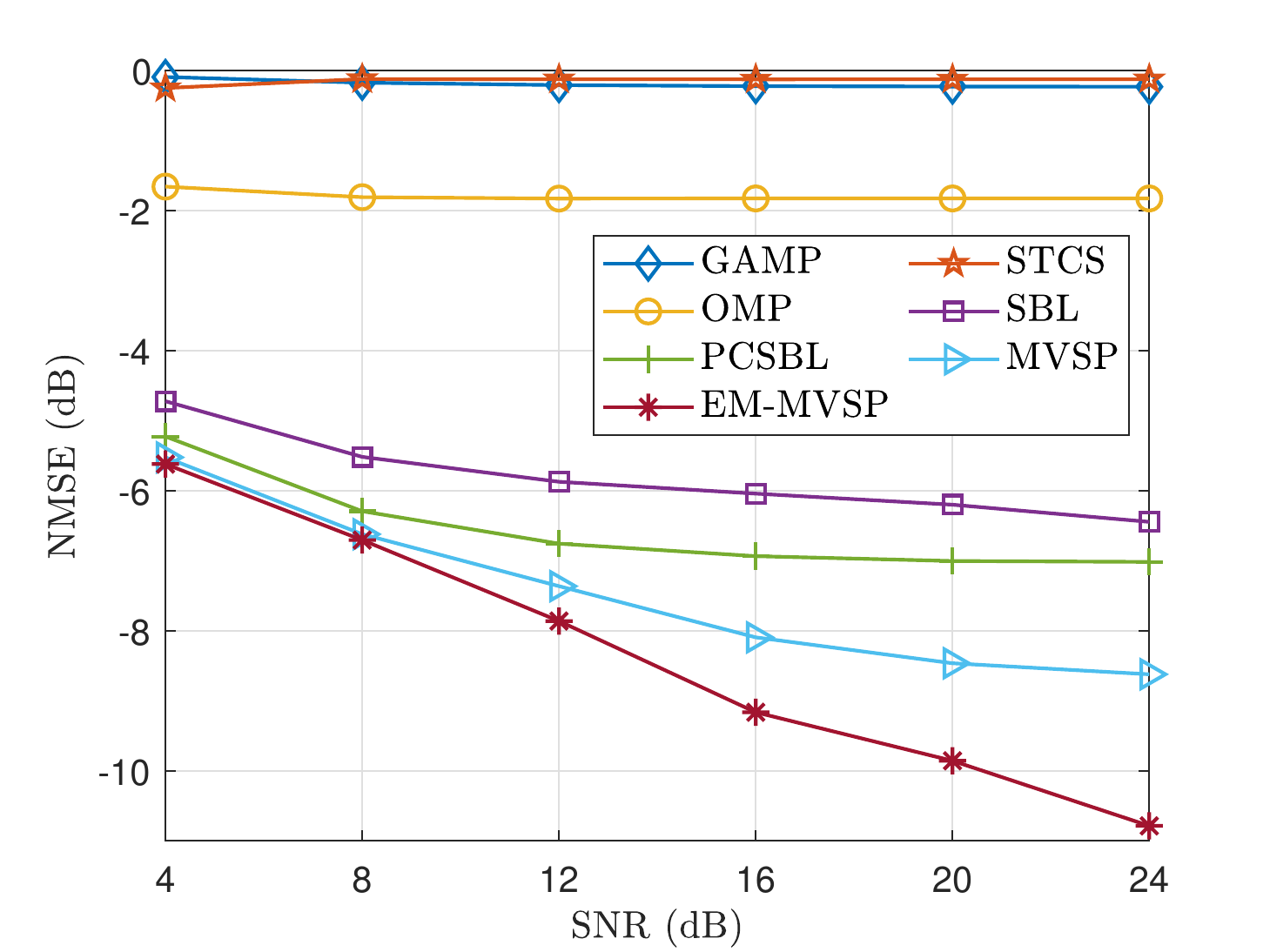}
\label{off-snr-nmse90}
\end{minipage}
}\subfigure[DAD  performance]{
\begin{minipage}[t]{0.5\linewidth}
\centering
\includegraphics[width=2.4in]{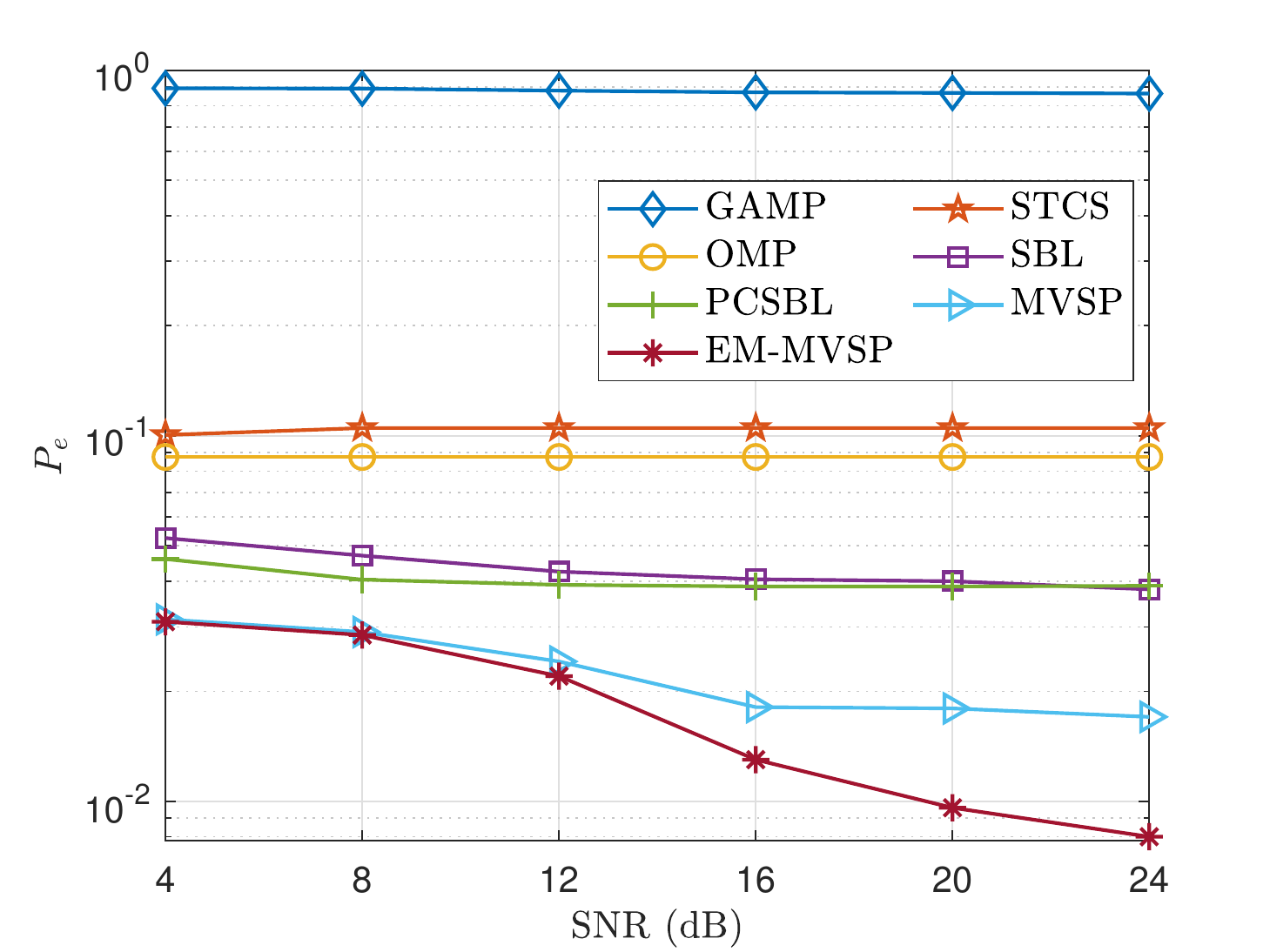}
\label{off-snr-perr90}
\end{minipage}
}  
\caption{CE and DAD performances under various SNR and off-grid settings, and $\alpha_e = 90^\circ$. Related parameters are $K=200$, $\rho = 0.1$, $N=4$,  $L=4$, $J=24$ and $M=32$. }
\label{off-90}
\end{figure}

%%% spar
\begin{figure}[t]
\centering
\subfigure[CE  performance]{
\begin{minipage}[t]{0.5\linewidth}
\centering
\includegraphics[width=2.4in]{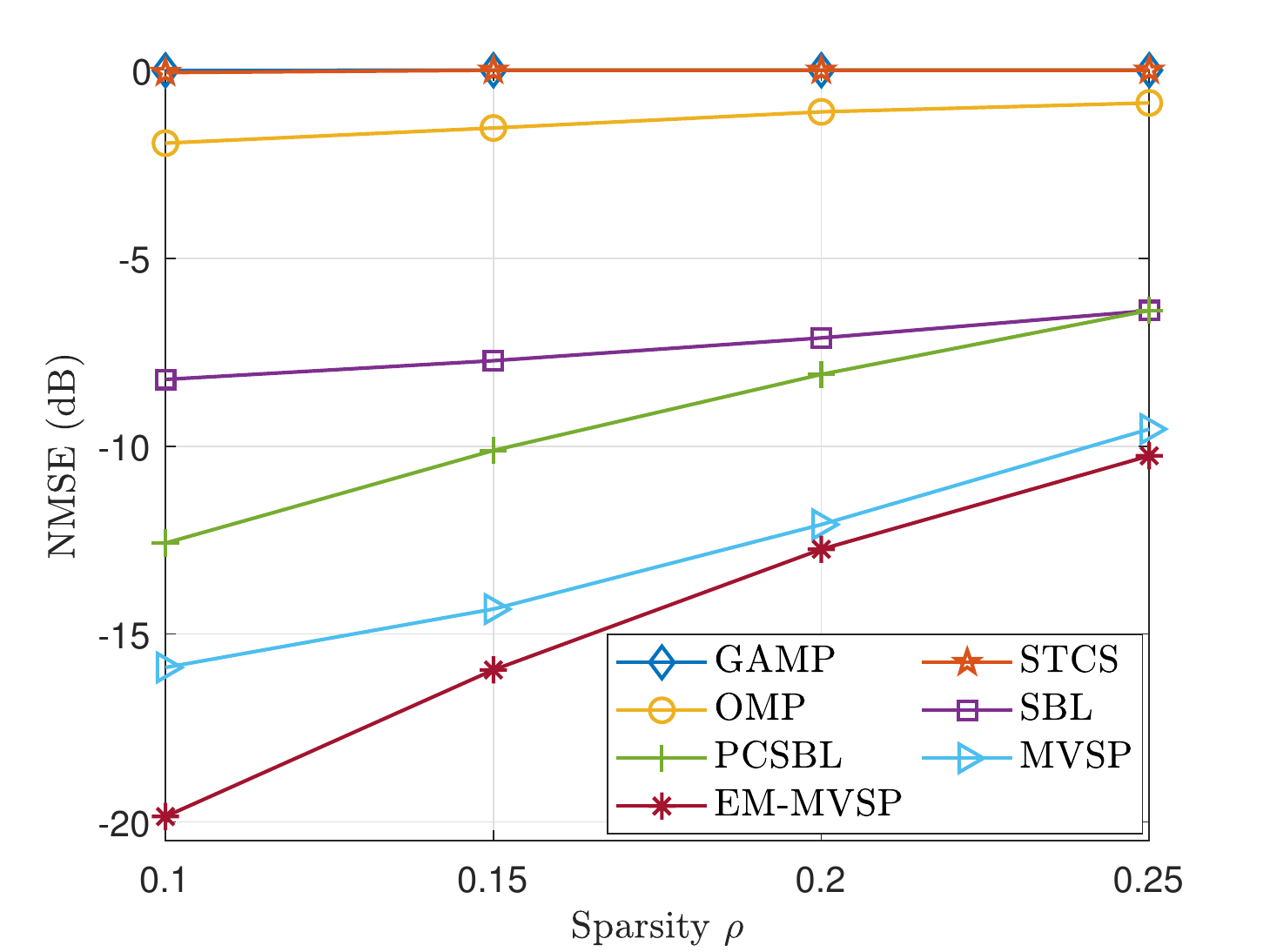}
\label{off-spar-nmse}
\end{minipage}
}\subfigure[DAD  performance]{
\begin{minipage}[t]{0.5\linewidth}
\centering
\includegraphics[width=2.4in]{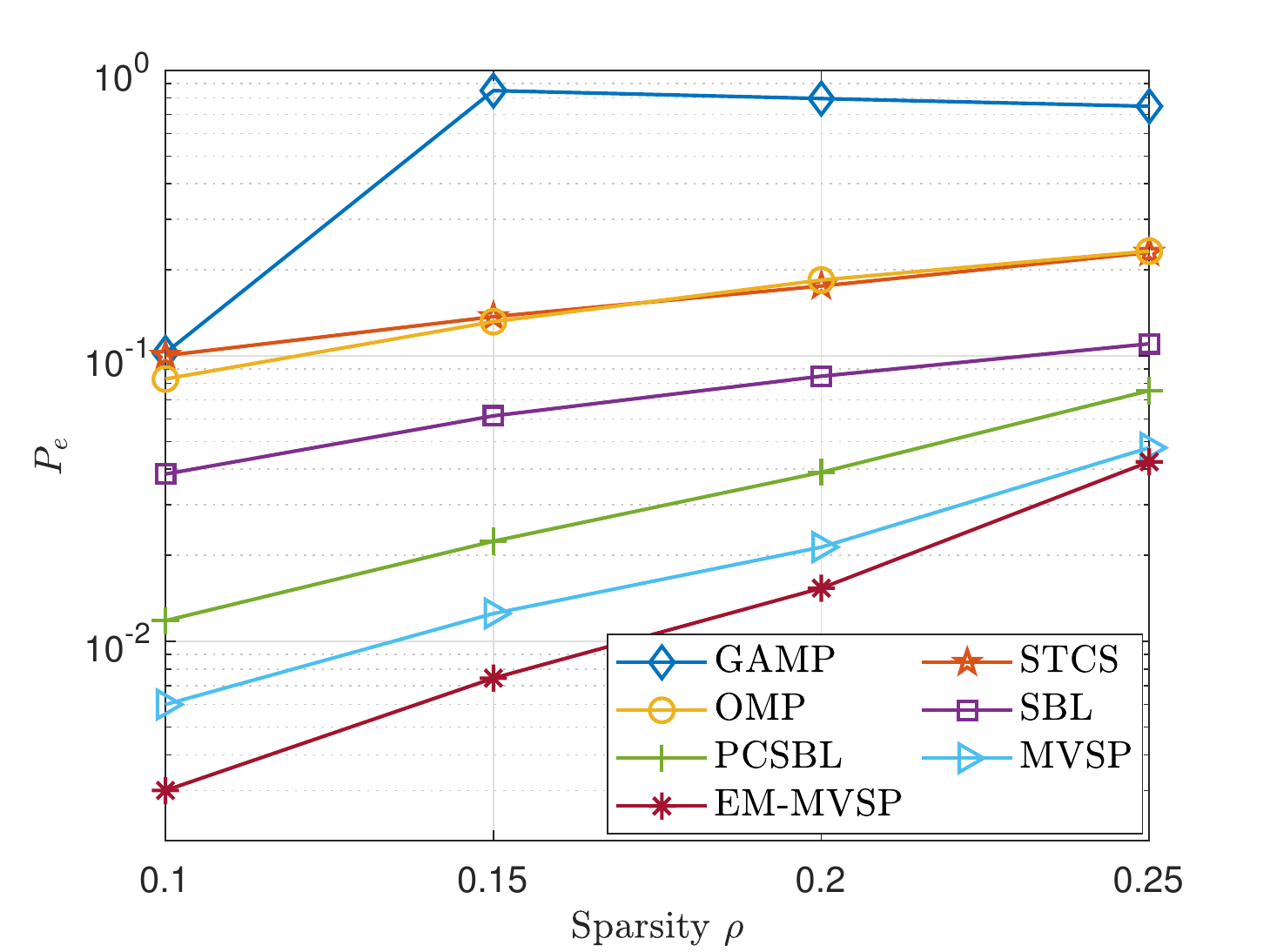}
\label{off-spar-perr}
\end{minipage}
}  
\caption{CE and DAD performances under various sparsity $\rho$ and off-grid settings. Related parameters are $K=200$, $N=4$,  $L=4$, $J=20$, SNR $=24$ dB, $M=32$ and $\alpha_e = 50^\circ$. }
\label{off-spar}
\end{figure}

We further consider the  mismatch between the real channel and the grid-based model, namely, the off-grid scenario. 
Fig. \ref{off-n-nmse}  shows the NMSE  against the number of repeated OFDM symbols $N$ in a super-symbol. As $N$ increases, it is interesting that different methods have different trends since more repeated OFDM symbols can improve the frequency resolution, but also results in the  correlation of measurement matrix.  The NMSEs of GAMP and OMP increase because they are sensitive to  measurement matrix, and repeated OFMD symbols  results in performance degradation.  As for PCSBL and SBL,   
% a slight performance gain is  obtained from frequency resolution improvement, and 
their NMSEs first slightly decrease and then increase. Then,  MVSP  are more robust to the measurement matrix, with  significant performance gain as $N$ increases. We see that  for MVSP more than 4 dB gain can be obtain from repeated OFDM symbols. When $N>4$, almost all the algorithms have a performance degradation due to the measurement matrix correlation.
In Fig. \ref{off-snr-nmse}, we show detection error probability   against $N$. The trade off between frequency resolution improvement and measurement matrix correlation is also obvious, which demonstrates the effectiveness of cascaded OFDM symbols.

Then we show the NMSE performance against the SNR with $\alpha_e= 50^\circ$ in Fig. \ref{off-snr-nmse}.  As the SNR increases, MVSP and EM-MVSP are at least 4 and 6 dB better than other baseline methods in NMSE, respectively. The proposed algorithms also behave well in DAD as shown in Fig. \ref{off-snr-perr}. We notice that PCSBL, MVSP and EM-MVSP have a significant performance gap compared with other methods in the considered SNR range, and their performance is similar at a lower SNR. This is because these three methods all consider the channel block-sparsity structure, but one-dimension block-sparsity in PCSBL and two-dimension block-sparsity in MVSP and EM-MVSP. In Fig. \ref{off-90}, we further consider the scenario with elevation angle $\alpha_e= 90^\circ$,  where the  Doppler effect is more severe. We see that MVSP and EM-MVSP still outperform the baselines. Compared with PCSBL, EM-MVSP have a performance gain of  more than 3 dB in NMSE and one order of magnitude in $P_e$. Thus, the proposed MVSP and EM-MVSP algorithms show performance advantages  under different Doppler effects.

Fig. \ref{off-spar} shows the CE and DAD performances of all the algorithms under varying sparsity $\rho$.  The proposed MVSP and EM-MVSP have a considerable performance gap within the considered sparsity range. Even at $\rho=0.2$, i.e., about 40 active devices access to the satellite, $P_e$ of EM-MVSP can reach  $10^{-2}$, which demonstrates the advantage of the proposed algorithms for  massive connectivity in satellite-IoT systems.

\section{Conclusion}
In this paper, we  studied the joint DAD and CE for GF-NORA in LEO satellite-IoT. We developed an OFDM-symbol  repetition technique to  better distinguish the Doppler shifts of the  LEO satellite channel.  We established a grid-based parametric system model, and showed that joint DAD and CE can be formulated as a CS problem. However, we pointed out that the measurement matrix of the problem exhibits special correlation structure, so that existing Bayesian CS algorithms such as AMP and Turbo-CS do not behave well. To address this issue, we proposed  the  MVSP algorithm which is robust to the sensing matrix and can efficiently exploit the channel sparsity in the delay-Doppler-user domain. 
We then used the EM method to learn the grid parameters and further improve the performance of MVSP. Simulation results demonstrated that the proposed algorithms significantly outperform the counterparts methods.
\bibliographystyle{IEEEtran}
\bibliography{myreference}

\end{document}